\documentstyle[epsfig]{mn}

\title{The outer-halo globular cluster NGC~6229.\\IV. Variable stars}

\author[J.~Borissova, M.~Catelan, T.~Valchev]
  {J.~Borissova,$^{1}$ M.~Catelan,$^{2}$\thanks{Hubble Fellow} 
   T.~Valchev$^{1}$\thanks{It is with deep sorrow that we announce 
   that T. Valchev passed away, at the age of 31, on December 
   31$^{\rm st}$, 1999, at the Pirin Mountain, Bulgaria.}\\
$^{1}$Institute of Astronomy, Bulgarian Academy of Sciences,
72~Tsarigradsko chauss\`ee, BG\,--\,1784 Sofia, Bulgaria \\
e-mail: jura@haemimont.bg \\
$^{2}$University of Virginia , Department of Astronomy,  
P.O. Box 3818, Charlottesville, VA 22903-0818 -- USA  \\
e-mail: catelan@virginia.edu\\
}

\date{Accepted 
      Received;
      in original }

\pubyear{2000}

\begin{document}

\maketitle

\label{firstpage}

\begin{abstract}
We present time-series $B$, $V$ photometry of NGC~6229, 
obtained with the purpose of providing the first extensive CCD
variability study of this cluster. As a result, we were able 
to obtain periods for all NGC~6229 variables, with the exception 
of five stars from the Borissova et al. (1997) candidate list 
located very close to the cluster center. Two variable stars 
suspected by Carney, Fullton, \& Trammell (1991) are 
first-overtone RR Lyrae (RRc) stars, whereas seven of the 12 
candidates from Borissova et al. (1997) are confirmed 
variables---three of them being fundamental RR Lyrae (RRab) 
pulsators, two first-overtone pulsators, one eclipsing binary 
(most likely an Algol system) and one bright star whose variability 
status could not be satisfactorily determined. 
A new image subtraction 
method (ISM) suggested by Alard (1999) has been employed which, 
together with the Welch \& Stetson (1993) technique, has allowed 
us to discover twelve new RR Lyrae variables in the cluster, for 
which ephemerides are provided. Ten of these are RRab's, whereas 
the other two are RRc's. As originally suggested by Mayer (1961), 
NGC~6229 is clearly an Oosterhoff type I globular cluster. 
We also confirm that V8 is a Population~II Cepheid of the 
W~Virginis type, and suspect that this is the case for V22 as 
well. 

The physical properties of the NGC~6229 RR Lyrae population are 
compared against those of M3 (NGC~5272) using several different 
methods, including a standard period-shift analysis. Possible 
differences between these two clusters are discussed. 

\end{abstract}


\section{Introduction}
In previous papers of this series, we have carried out extensive 
analyses of the colour-magnitude 
diagram (CMD) of the outer-halo globular cluster (GC) 
NGC~6229 (C1645+476). Borissova et al. (1997, hereafter BCSS97) 
presented the first CMD for the crowded core of the cluster, 
finding the horizontal-branch (HB) morphology of NGC~6229 to be 
bimodal, containing a prominent blue tail with at least 
one `gap'. Catelan et al. (1998) carried out a detailed 
theoretical and statistical analysis of HB bimodality and `gaps' 
(in NGC~6229 and other GCs). Borissova et al. (1999) obtained 
the first deep CMD of NGC~6229 clearly identifying the 
main-sequence turnoff of the cluster, and found its metallicity 
and age to be very similar to those of M5 (NGC~5904); the 
possible presence of `extreme' HB stars was also highlighted. 

The purpose of the present paper, the fourth in our  
NGC~6229 series, is to provide the first detailed CCD 
investigation of the variable star population in this very 
distant GC. 

Using unpublished data obtained by H.~Shapley, Davis (1917) 
reported on the discovery of the first variable star in 
NGC~6229---the variable currently known as V8. She called 
attention to the brightness of the variable in comparison 
with other non-variable stars in the cluster, claiming it to 
be the second brightest NGC~6229 star when at its maximum. 
Indeed, Baade (1945) would later rediscover this star and 
assign it a `long-period Cepheid' status. 
Baade's was the first extensive variability study 
to be published for NGC~6229. It was based on 46 
photographic plates obtained in the years 1932-1935 at the 
Mount Wilson Observatory. By intercomparison of 11 pairs of 
plates with a blink-comparator, he discovered 20 new variable 
stars, all `of the cluster type' (i.e., RR Lyrae 
stars). Sawyer (1953) later extended the list of known variable 
stars in NGC~6229 to 22 with the report of a (bright) variable 
star (V22) very close to the cluster core. 

None of these papers 
reported periods for the variables. Mayer (1961) used Baade's 
(1945) data to provide periods for 10 of the RR Lyrae variables. 
Mannino (1960) published extensive photometry for 16 among the 
22 known NGC~6229 variables---and, using both his 92 plates 
(from Asiago, obtained over the years 1956-1957-1958) and 
Baade's 46 plates, provided periods for 12 of these stars, 
assuming that the periods stayed constant between the epochs 
of his and Baade's observations. We note that, as a rule, in 
Sawyer Hogg's (1973) catalogue Mannino's periods were adopted, 
those from Mayer being used only for variables V6, V7, and 
V17, for which Mannino did not give periods: for V6 and V7, 
Mannino did not obtain any new data, whereas for V17, and 
also V3, 
V4, and V21, he was simply unable to derive representative 
periods. Therefore, Sawyer Hogg's catalogue, and also its 
recent update (Clement 1997), contain periods for all 22 
variables except V3, V4, V11, V12, V18, V21, and V22. 

More recently, Carney, Fullton, \& Trammell (1991, hereafter 
CFT91), in their CCD analysis of the CMD of NGC~6229, provided 
single-frame magnitudes and colours for most of Baade's (1945) 
variables. In addition to this, they identified two additional 
stars as likely variables (see also Searle \& Zinn 1978), 
besides four other stars for which there was some evidence for 
photometric variability. Finally, BCSS97, using the Kadla \& 
Gerashchenko (1982) method and a wavelet transform technique 
to allow CCD photometry in the very dense core of this GC, 
provided a list containing an additional 12 RR Lyrae variable 
candidates. 

In the present paper, we will provide the first CCD light curves 
for all previously known or suspected/confirmed variables in 
NGC~6229.\footnote{For V3, which is outside our observed field, 
we shall determine the period and variability class on the basis 
of the data published by Mannino (1960).} Besides, using a new 
and very powerful `image subtraction method' (ISM) developed by 
Alard (1999)---the {\sc isis} package---we will also provide 
light curves for twelve new variables discovered in the course 
of the present investigation.   

We begin in the next section by describing our observations and 
data reduction techniques. In Section~3, the light curves and 
the variables' main parameters (periods, colours, magnitudes, 
amplitudes), as obtained using both {\sc daophot} and {\sc isis}, 
are provided. In Section~4, Blazhko stars are identified. In 
Section~5, the Bailey (period-amplitude) diagram is 
discussed. In Section~6, Fourier decomposition of the light 
curves is used to derive the stellar physical parameters, 
following methods developed by Simon and Clement (RRc stars) 
and by Kov\'{a}cs and Jurcsik (RRab stars). In Section~7, the 
nature of a few particularly noteworthy cluster variables is 
discussed. Finally, our concluding remarks are presented in 
Section~8.

\section{Observations and data reduction}

\subsection{Observations}
Time-series photometry was obtained during the interval 1997 June 28 -
1999 September 26 with the 2m Ritchey-Chr\'etien telescope of the 
Bulgarian National Astronomical Observatory `Rozhen' with a 
Photometrics $1024\times 1024$ CCD camera.  The scale at the
Cassegrain-focus CCD was $0.33\arcsec\,{\rm pixel}^{-1}$ and the
observing area was $5.6\arcmin \times\ 5.6\arcmin$, centered on the
cluster centre. 

Approximately 200 $B$ and $V$ frames were obtained, with exposure 
times ranging from 120 to 300~sec.  The seeing during these 
observations was between $\approx 1\arcsec$ and $1.6\arcsec$.

\subsection{Data reduction: {\sc daophot}}

All frames were reduced using {\sc daophot} (Stetson 1993) and 
transformed to the standard system independently. In the cases when 
the seeing was not stable the data were discarded. The data 
reduction and absolute photometric calibration of the frames are the 
same as those obtained and used for the construction of the deep
CMD of NGC~6229 (Borissova et al. 1999). The internal error of each 
measurement for the stars in NGC~6229 depends on the seeing and 
crowding conditions and is on average $\approx 0.02$~mag for 
$V < 21$~mag and $B < 21$~mag. For more details on the error and 
completeness analysis, see BCSS97 and Borissova et al. (1999). 
A comparison with the photometry of CFT91 for the non-variable stars 
in the magnitude interval $B,\,V  = 15 - 20$~mag shows that the   
two datasets are in good agreement, although in some frames we cannot 
exclude systematic zero point differences, though not larger than 
0.02--0.03~mag. When possible, datapoints from the photometry lists 
of CFT91 and BCSS97 were included in the present investigation.

In order to search for new variables in the studied field of NGC~6229, 
we applied the Welch \& Stetson (1993) method. Since we have 
observations made nearly simultaneously in two passbands ($B$ and $V$) 
and in several different epochs separated at least by a month, we can 
calculate the Welch \& Stetson variability index `$I$' for all stars. 
As a result, two previously unknown variable stars---V37 and V38---were 
detected. They are both fundamental RR Lyrae pulsators (see below). 

\subsection{Data reduction: {\sc isis}}

Alard (1999) and Alard \& Lupton (1998) presented a new ISM 
(`{\sc isis}') aimed at obtaining better light curves in crowded 
fields such as GC cores. The capabilities of the method have recently 
been demonstrated by Olech et al. (1999) in a study of variable stars 
in the core of M5.  

In order to improve the quality of the light curves of the known 
variables in NGC~6229 and to search for new variables, all our CCD 
frames were re-reduced using Alard's (1999) code. Ours is one of 
the very first applications of the method in a study of stellar 
variability in GCs.  

Twenty-nine of our RR Lyrae stars have magnitudes measured using 
both the standard {\sc daophot} package (which places them in the 
standard Johnson system) and the non-standard {\sc isis} package 
(which is only capable of providing relative fluxes, measured in 
ADU). We have used the $B$ and $V$ magnitudes for such stars in 
order to try to obtain transformation equations linking standard 
magnitudes and {\sc isis} fluxes. As far as we are aware, such a 
discussion has not been presented in the literature so far, the 
{\sc isis} results usually being reported in ADU only (see, e.g., 
Olech et al. 1999 in the case of M5). 

Unfortunately, for most of our stars, the transformations appear 
to be affected by some non-linearity. More specifically, the 
brightest point in the light curves, as obtained in the standard 
system, usually appears fainter than that calculated from the ISM 
fluxes. If, as a first approximation, we use a linear transformation 
of the form  
$m_{\rm ISM} = \alpha + \beta \times m_{\rm ISM}^{\rm ADU}$ 
(where $m$ represents ISM magnitudes in $B$ or $V$), the $\beta$ 
coefficient turns out to be 0.00007 and 0.00014 for the $V$ and $B$ 
filters, respectively, and appears to be fairly uniform for our  
RR Lyrae sample. The zero-point coefficients $\alpha$, however, 
do seem to vary slightly from star to star, so that we cannot  
convert the ISM fluxes into their Johnson-equivalent magnitudes in 
an entirely consistent way. In the present paper, for the RR Lyrae 
stars which were measured only with {\sc isis}, approximate 
ISM-based $B$ and $V$ magnitudes were obtained using the $\beta$ 
value stated above; the adopted $\alpha$ was taken from the 
standard magnitude measured in the reference frames. 

These problems notwithstanding, our analysis clearly indicates that 
{\sc isis} is an extremely powerful tool to detect new variables 
and determine their periods, particularly in very crowded fields. 
Indeed, applying the method to NGC~6229, all known 
variables in our observed field (which excludes only V3) were 
rediscovered, and light curves of as many 
as ten previously unknown variables were obtained. We label 
these new discoveries V39--V48 in what follows.

\section{Periods and light curves}

\begin{figure*}
 \centerline{\epsfig{figure=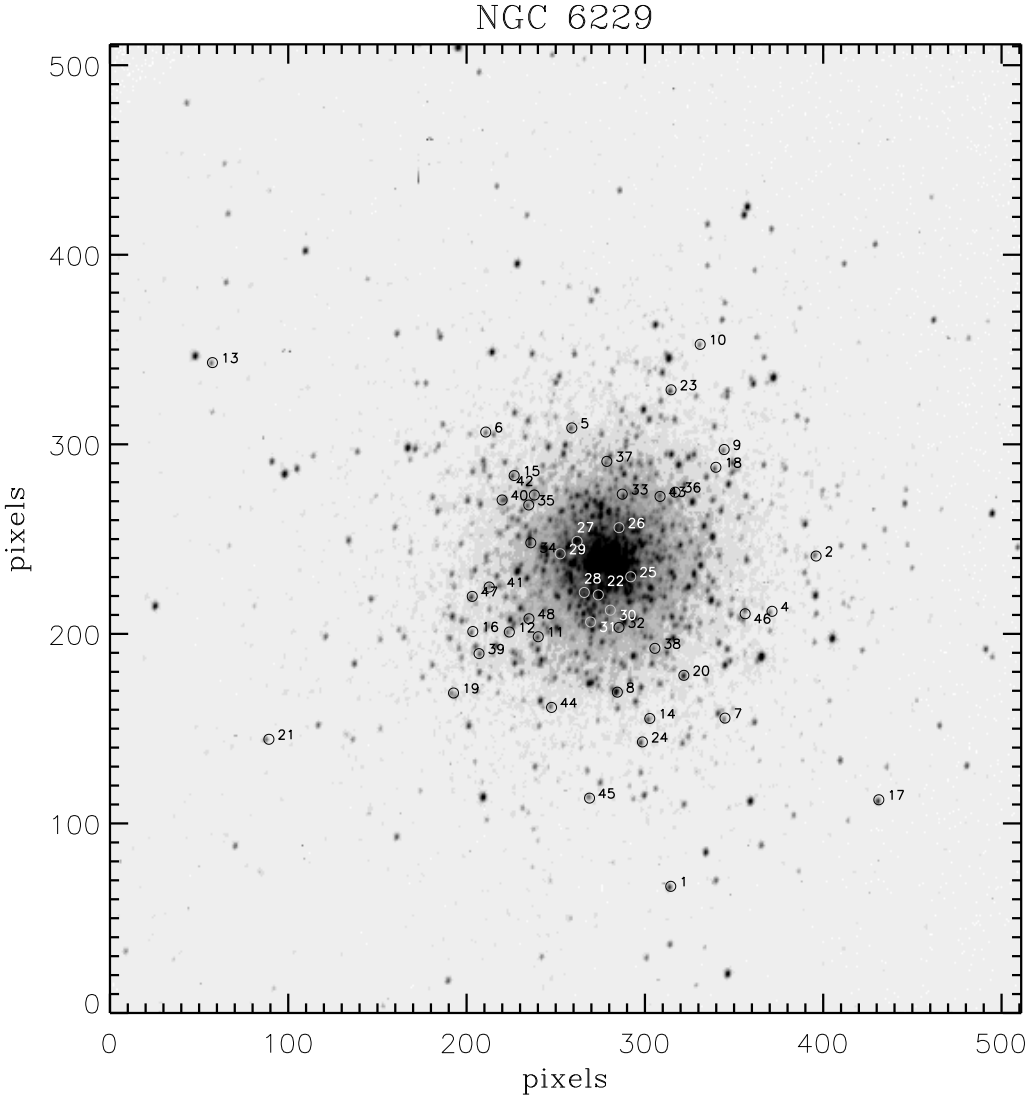}}
      \caption{Finding chart for all known variables in NGC~6229 in 
      a $5.6\arcmin \times\ 5.6\arcmin$ area, centered on the cluster 
      centre. North is up and east to the left. V3 is outside our 
      field of view. 
}
      \label{Fig06}
\end{figure*}

Our `target list' contains 50 variable and possible variable stars
based on Baade (1945), Sawyer (1953), CFT91, BCSS97 and the present 
work (see Sect.~2). 
We have reanalysed the periods of all Baade stars and determined  
the periods of suspected ones using a least-squares periodogram 
analysis by means of the phase dispersion minimization (PDM) task 
available in {\sc iraf} and a period-finding program based on 
Lafler \& Kinman's (1965) `theta' statistic. We obtained new 
light curves and re-derived the periods for 20 variable stars 
listed in Sawyer Hogg's (1973) catalogue, and also the first 
light curves for RR Lyrae candidates No.~155 and No.~88 from  
CFT91 and 3, 7, 8, 9, 10, 11 and 12 from BCSS97. 
We have not detected any significant variability (at a level  
$\gid 3 \sigma$) for the stars No.~4, No.~105 and No.~134 
suspected by CFT91. A finding chart for the NGC~6229 variables 
is provided in Fig.~\ref{Fig06}. 

\begin{figure*}
 \centerline{\epsfig{figure=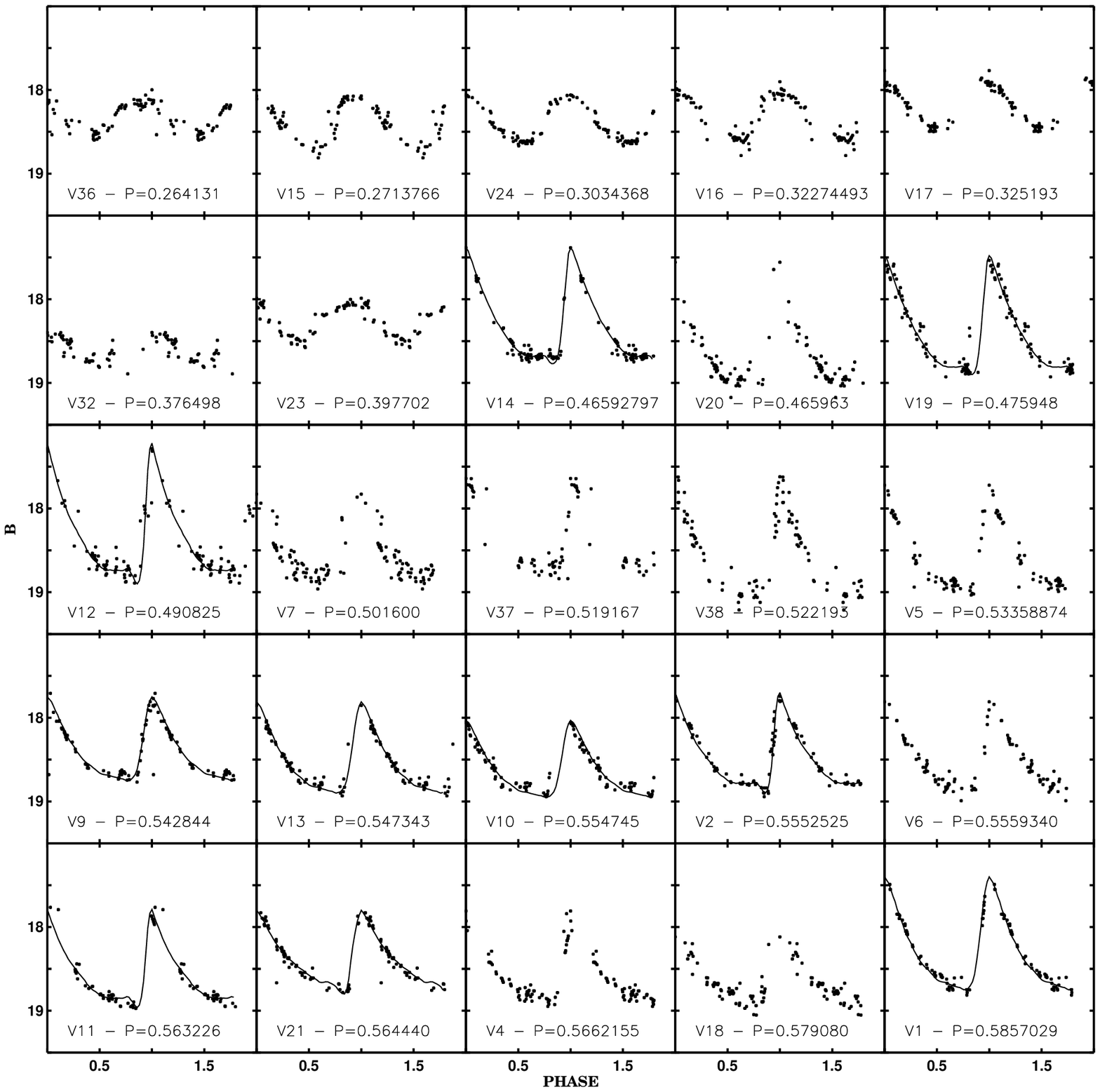, height=5.65in, width=5.75in}}
 \vspace{0.25 in}
      \caption{$B$ light curves for NGC~6229 variables, as 
      obtained using {\sc daophot}. The `best-fitting' fiducial 
      light curves, as obtained from Layden (1998), are overplotted 
      on the data for the `regular' variables (see text).}
      \label{Fig01}
\end{figure*}

\begin{figure*}
 \centerline{\epsfig{figure=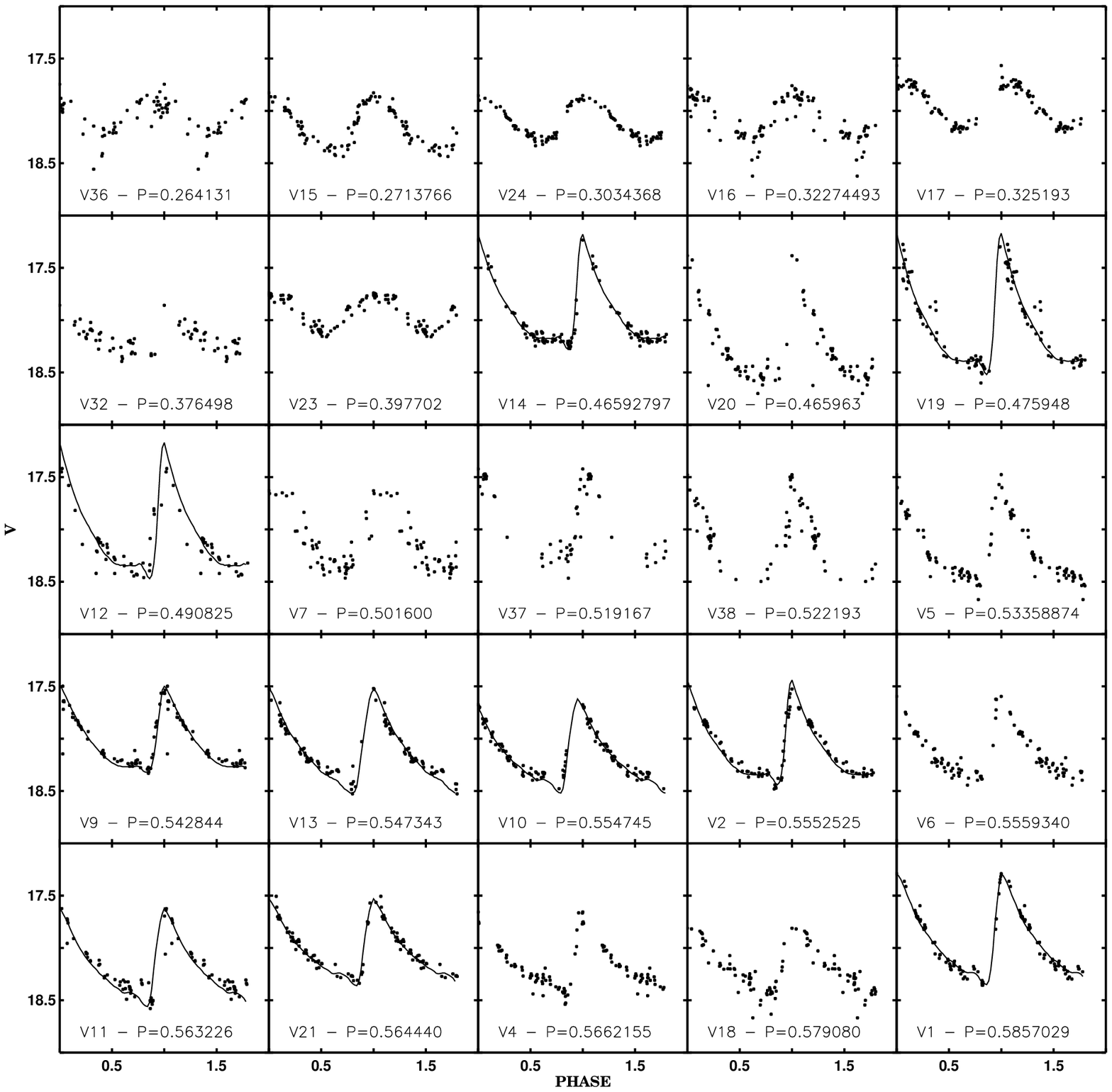, height=5.65in, width=5.75in}}
 \vspace{0.25 in}
      \caption{$V$ light curves for NGC~6229 variables, as 
      obtained using {\sc daophot}. The `best-fitting' fiducial 
      light curves, as obtained from Layden (1998), are overplotted 
      on the data for the `regular' variables (see text).}
      \label{Fig02}
\end{figure*}

The $B$ and $V$ RR Lyrae light curves obtained with {\sc daophot} 
are displayed in Fig.~\ref{Fig01} and Fig.~\ref{Fig02}, 
respectively.

\begin{figure*}
 \centerline{\epsfig{figure=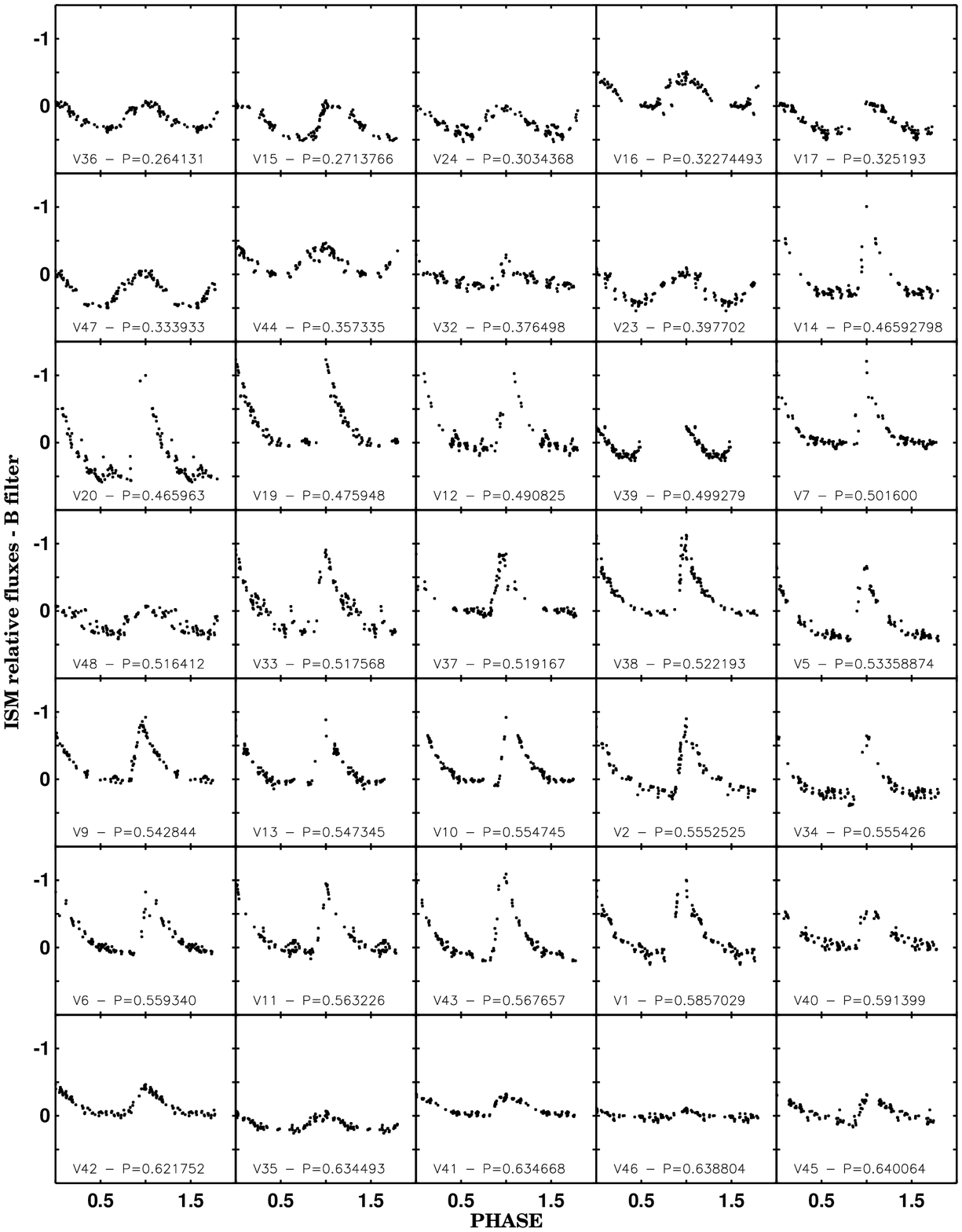, height=7.25in, width=5.75in}}
 \vspace{0.25 in}
      \caption{ISM relative fluxes in $B$, as derived using
      {\sc isis}, as a function of 
      phase for the NGC~6229 variables.}
      \label{Fig03}
\end{figure*}

\begin{figure*}
 \centerline{\epsfig{figure=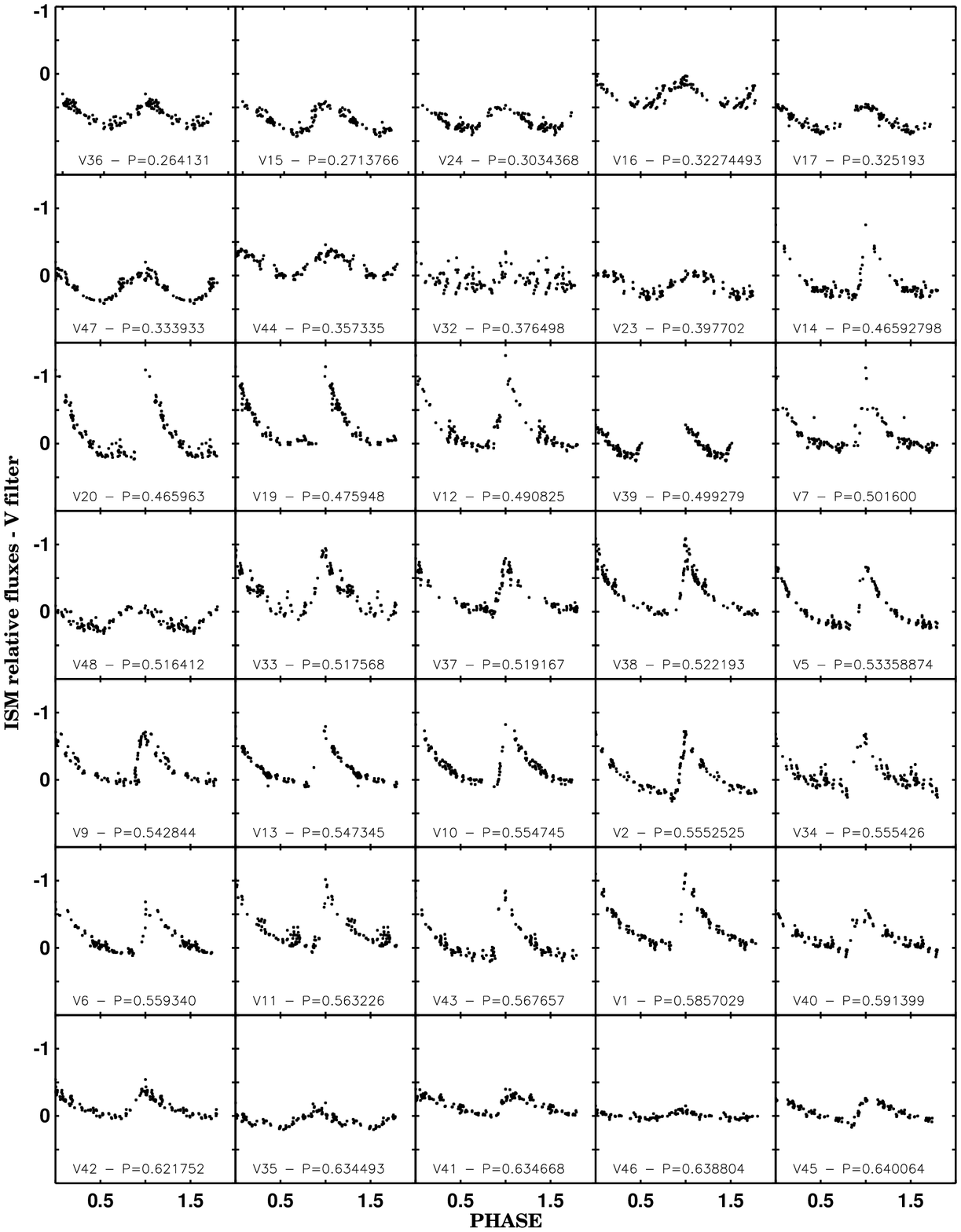, height=7.25in, width=5.75in}}
 \vspace{0.25 in}
      \caption{ISM relative fluxes in $V$, as derived using
      {\sc isis}, as a function of 
      phase for the NGC~6229 variables.}
      \label{Fig04}
\end{figure*}

\begin{table*}
 \begin{minipage}{160mm}
 \caption{Periods and mean magnitudes for variables from Sawyer Hogg (1973)}
 \label{Tab01}
 \begin{tabular} {@{}llrrrccccccc@{}}

Name&Type&$X$&$Y$&Period&$\langle B\rangle$&$\langle V\rangle$&$\langle B-V\rangle$&
         $\langle B \rangle - \langle V \rangle$&$(B-V)_{\rm mag}$ & $A_{B}$&$A_{V}$ \\
           & & ($\arcsec$) & ($\arcsec$)  &    (d) &&&&&&\\

V1    & RRab     & $-24.6$  & $-105.5$ &   0.5857029  & 18.23 & 17.89 & 0.36 & 0.34 & 0.38& 1.35 & 1.06\\
V2    & RRab     & $-71.9$  & $+4.9$   &   0.5552525  & 18.41 & 18.06 & 0.38 & 0.35 & 0.38& 1.20 & 1.00\\
V3    & RRab\footnote{Data from Mannino (1960).}     
                 & $-195.7$ & $41.3$   &   0.575218   & $-$   & $-$   & $-$  & $-$  & $-$ & $-$   & $-$\\
V4    & RRab\footnote{Blazhko variable (see text).} 
                 & $-56.8$  & $-14.3$  &   0.5662155  & 18.44 & 18.09 & 0.38 & 0.35 &0.39 & 1.15 & 0.75\\
V5    & RRab$^b$ & $+14.5$  & $+44.1$  &   0.53358874 & 18.51 & 18.14 & 0.43 & 0.37 &0.40 & 1.30 & 0.95\\
V6    & RRab$^b$ & $+44.1$  & $+41.5$  &   0.559340   & 18.46 & 18.09 & 0.44 & 0.37 &0.41 & 1.15 & 0.82\\
V7    & RRab$^b$ & $-41.7$  & $-49.9$  &   0.501600   & 18.42 & 18.05 & 0.42 & 0.37 &0.44 & 1.13 & 1.05\\
V8    & W Vir    & $-4.1$   & $-42.1$  &  14.840457   & 16.29 & 15.53 & 0.79 & 0.76 &0.78 & 1.05 & 0.77\\
V9    & RRab     & $-38.9$  & $+38.3$  &   0.542844   & 18.36 & 17.93 & 0.40 & 0.43 &0.39 & 1.00 & 0.82\\
V10   & RRab     & $-29.5$  & $+72.7$  &   0.554745   & 18.56 & 18.16 & 0.43 & 0.40 &0.44 & 0.92 & 0.90\\
V11   & RRab     & $+23.9$  & $-25.0$  &   0.563226   & 18.57 & 18.16 & 0.43 & 0.41 & 0.43& 1.17 & 0.94\\
V12   & RRab     & $+34.5$  & $-23.7$  &   0.490825   & 18.37 & 18.01 & 0.36 & 0.36 &0.35 & 1.68 & 1.30\\
V13   & RRab     & $+140.2$ & $+61.3$  &   0.547345   & 18.47 & 18.07 & 0.42 & 0.40 &0.44 & 1.09 & 0.99\\
V14   & RRab     & $-15.5$  & $-50.7$  &   0.46592798 & 18.27 & 17.91 & 0.43 & 0.36 &0.43 & 1.38 & 1.10\\
V15   & RRc      & $+34.2$  & $+27.5$  &   0.2713766  & 18.37 & 18.14 & 0.25 & 0.23 &0.28 & 0.67 & 0.56\\
V16   & RRc      & $+47.0$  & $-24.2$  &   0.32274493 & 18.32 & 18.06 & 0.28 & 0.26 &0.23 & 0.63 & 0.48\\
V17   & RRc      & $-96.3$  & $-75.0$  &   0.325193   & 18.16 & 17.94 & 0.22 & 0.22 & 0.22& 0.61 & 0.47\\
V18   & RRab$^b$ & $-36.1$  & $+32.2$  &   0.579080   & 18.55 & 18.15 & 0.43 & 0.40 &0.42 & 0.90 & 0.65\\
V19   & RRab     & $+53.4$  & $-44.4$  &   0.475948   & 18.30 & 17.97 & 0.37 & 0.33 &0.36 & 1.42 & 1.35\\
V20   & RRab$^b$ & $-27.5$  & $-36.1$  &   0.465963   & 18.53 & 18.16 & 0.40 & 0.37 &0.41 & 1.60 & 1.30\\
V21   & RRab     & $+117.3$ & $-61.6$  &   0.564440   & 18.38 & 17.99 & 0.39 & 0.39 &0.46 & 1.00 & 0.83\\
V22   & W Vir?\footnote{See Sect.~7.2.}
                 & $+3.8$   & $-10.5$  &   15.8373    & $-$   & $-$   & $-$  & $-$  & $-$ & $-$   & $-$\\
\end{tabular}
\end{minipage}
\end{table*}

Using the same statistical approach---the PDM task and 
Lafler-Kinman's `theta' statistic---we analyzed relative ISM 
fluxes of all variables. In general, we recovered the same 
periods obtained with {\sc daophot}, but in some cases we 
were able to improve the derived periods significantly. 
The light curves in ISM relative fluxes, as obtained from 
the $B$ and $V$ data for the RR Lyrae stars using {\sc isis}, 
are displayed in  Fig.~\ref{Fig03} and Fig.~\ref{Fig04}, 
respectively.

The quality of the light curves, particularly for the variable 
stars in the inner regions of NGC~6229, is noticeably better  
in the {\sc isis} than in the {\sc daophot} case (compare, e.g., 
V36 in Fig.~\ref{Fig01} and Fig.~\ref{Fig03}). However, even 
with {\sc isis} we could not obtain representative light curves 
for candidates numbered 1, 2, 4, 5 and 6 by BCSS97. All these 
stars are less than 0.3~arcmin from the cluster centre (see 
Fig.~\ref{Fig06}) and cannot be resolved on our moderate 
quality CCD material for this very distant and fairly dense GC. 
However, with the exception of BCSS97--1, a few of our best 
ISM-subtracted $B$ images {\em do} show signs of photometric 
variability for these stars, which thus remain likely variable 
candidates. Follow-up studies are clearly needed to verify 
these stars' variability status.

\begin{table*}
\begin{minipage}{175mm}
\caption {Periods and mean magnitudes for variables suspected by CFT91}
\label{Tab02}
\begin{tabular} {lllrrrccccccc}
Name&CFT91 &Type&$X$&$Y$&Period&$\langle B\rangle$&$\langle V\rangle$&$\langle B-V\rangle$&
         $\langle B \rangle - \langle V \rangle$&$(B-V)_{\rm mag}$ & $A_{B}$&$A_{V}$ \\
             & name    & & ($\arcsec$) &   ($\arcsec$)  &    (d)&&&&&&\\
V23 & No.~88  & RRc & $-19.8$ & $+57.8$ & 0.397702  & 18.23 & 17.91 & 0.33 & 0.32 & 0.31& 0.47 & 0.42\\
V24 & No.~155 & RRc & $-12.9$ & $-58.3$ & 0.3034368 & 18.33 & 18.07 & 0.27 & 0.26 &0.27 & 0.60 & 0.46\\
\end{tabular}
\end{minipage}
\end{table*}

\begin{figure*}
 \centerline{\epsfig{figure=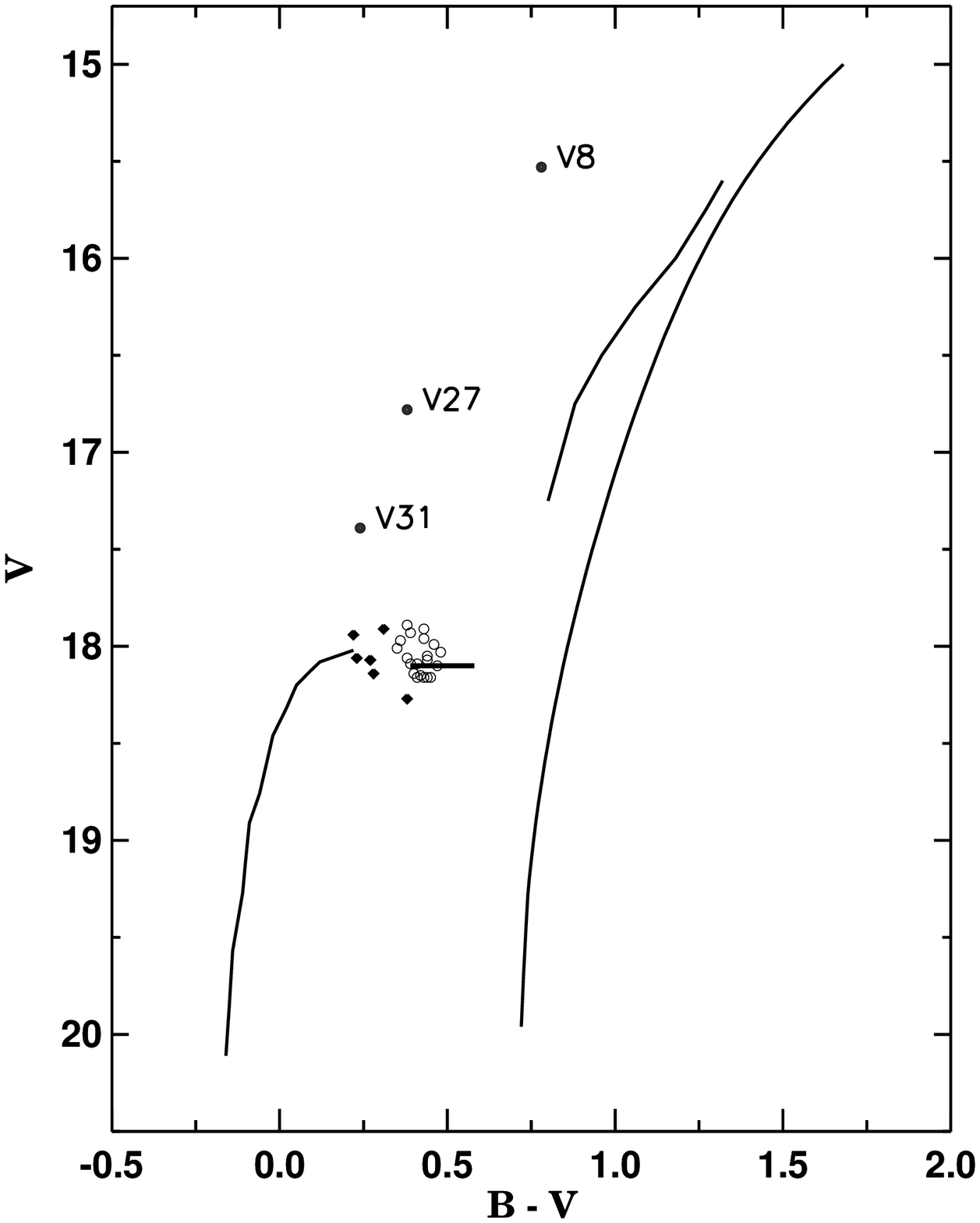, height=4.75in, width=10.5cm}}
      \caption{Detailed plot showing the RRc (filled squares), 
      RRab (open circles), a Pop.~II Cepheid (V8) and two other 
      variables of uncertain variability status (filled circles)  
      in the colour-magnitude diagram. 
      The fiducial lines of the cluster (from Borissova et al. 
      1999) are also overplotted.
       }
      \label{Fig05}
\end{figure*}

In Figs.~\ref{Fig01}--\ref{Fig04}, 
the variable stars are plotted in order of increasing period to 
illustrate the changes in the light curves with period. The 
difference in light curve shape for stars with periods 
greater than $0\fd40$ compared with those with shorter period 
is obvious. Nine variables (V15, V16, V17, V23, V24, V32, V36, 
V44, V47) can be classified as RRc stars (i.e., first overtone 
pulsators). The light curves of the remaining 29 variables are 
more asymmetric and have larger amplitudes, and these are 
undoubtedly RRab Lyrae stars.  

\begin{table*}
\begin{minipage}{175mm}
\caption {Periods and mean magnitudes for variables suspected by BCSS97}
\label{Tab03}
\begin{tabular} {@{}lllrrrccccccc@{}}
Name&BCSS97 &Type&$X$&$Y$&Period&$\langle B\rangle$&$\langle V\rangle$&$\langle B-V\rangle$&
         $\langle B \rangle - \langle V \rangle$&$(B-V)_{\rm mag}$ & $A_{B}$&$A_{V}$ \\
             & name    & & ($\arcsec$) &   ($\arcsec$)  &    (d)&&&&&&\\
V27 &  3 & ?        & $+11.7$  & $+6.9$  &  1.13827 & 17.13 & 16.78 & 0.37 & 0.35 &0.38 & 0.82 & 0.79\\
V31 &  7 & Algol?   & $+6.2$  & $-19.5$ &  0.698893 & 17.62 & 17.39 & 0.26 & 0.23 &0.24 & 1.14 & 1.00\\
V32 &  8 & RRc\footnote{Field variable? See Sect.~7.1.}      
                    & $-3.8$  & $-20.9$ &  0.376498 & 18.58 & 18.27 & 0.34 & 0.31 &0.38 & 0.45 & 0.43\\
V33 &  9 & RRab\footnote{Blazhko variable (see text).} 
                    & $-3.7$  & $+22.9$ &  0.517568 & 18.52 & 18.03 & 0.43 & 0.49 &0.48 & 1.29 & 1.06\\
V34 & 10 & RRab$^b$ & $+27.8$ & $+5.8$  &  0.555426 & 18.24 & 17.92 & 0.32 & 0.32 &0.29 & 1.04 & 0.94\\
V35 & 11 & RRab$^b$ & $+28.9$ & $+18.2$ &  0.634493 & 18.60 & 18.16 & 0.45 & 0.44 & 0.45& 0.40 & 0.32\\
V36 & 12 & RRc      & $-22.1$ & $+24.2$ &  0.264131 & 18.33 & 18.07 & 0.26 & 0.26 & 0.27& 0.47 & 0.40\\
\end{tabular}
\end{minipage}
\end{table*}

The adopted ephemerides are listed in 
Tables~\ref{Tab01},~\ref{Tab02},~\ref{Tab03} and~\ref{Tab04}. 

Table~\ref{Tab01} contains the list of variable stars discovered by 
Davis (1917), Baade (1945), and Sawyer (1953). 
Column~(1) lists the star number from Sawyer Hogg's 
(1973) catalogue, and column~(2) gives the variability type of the star. 
Stars identified as Blazhko variables are noted. In columns~(3) and 
(4), we give the $X$ and $Y$ coordinates (in arcsec) in the 
Sawyer Hogg (1973) system. Column~(5) provides the period in days.  
Columns~(6) and (7) give the intensity-mean magnitudes in $B$ and 
$V$, respectively; those values were obtained by directly averaging
over the pulsation cycle in intensity. Likewise, columns~(8) and (9) 
provide the intensity-mean colours $\langle B-V \rangle$ (obtained 
by directly averaging over the colour curves in intensity units) and 
$\langle B \rangle - \langle V \rangle$ [simply obtained by 
subtracting the values in column~(7) from those in column~(6)]. 
Column~(10) gives the magnitude-weighted mean color, 
$(B-V)_{\rm mag}$, of the stars. Finally, columns~(11) and (12) 
provide the estimated amplitudes in $B$ and $V$, respectively.  

In order to derive the mean magnitudes and colours, the $B$ and 
$V$ light curves were fitted by smooth curves (e.g., Carretta et al. 
1998). The smoothing was performed on the light curves in intensity 
units using polynomials of various orders. For each star, the ISM 
light curves were also taken in consideration, but {\em only} in 
the sense that all points in the $B$ and $V$ light curves which 
clearly deviate from their corresponding ones in the ISM light 
curves were simply rejected. 

\begin{table*}
 \begin{minipage}{160mm}
 \caption {Periods and mean magnitudes for variables discovered in the present investigation}
 \label{Tab04}
 \begin{tabular} {@{}llrrrccccccc@{}}

Name&Type&$X$&$Y$&Period&$\langle B\rangle$&$\langle V\rangle$&$\langle B-V\rangle$&
         $\langle B \rangle - \langle V \rangle$&$(B-V)_{\rm mag}$ & $A_{B}$&$A_{V}$ \\
           & & ($\arcsec$) & ($\arcsec$)  &    (d) &&&&&&\\

V37 & RRab\footnote{Blazhko variable (see text).}
               & $+2.0$  & $+33.5$ & 0.519216 & 18.32 & 17.96 & 0.38 & 0.36  &0.43    & 1.17  & 0.92 \\
V38 & RRab$^a$ & $-16.5$ & $-27.4$ & 0.522193 & 18.52 & 18.10 & 0.43 & 0.42  &0.47    & 1.55  & 1.05 \\
V39 & RRab     & $+44.9$ & $-31.2$ & 0.499279 & 18.4: & 18.2: & $-$  & 0.2: & $-$     & 0.8: & 0.8:\\
V40 & RRab$^a$     & $+38.2$ & $+19.6$ & 0.591399 & 18.7: & 18.3: & $-$  & 0.4: & $-$ & 0.6: & 0.6:\\
V41 & RRab$^a$     & $+41.9$ & $-9.2$  & 0.634668 & 18.7: & 18.3: & $-$  & 0.4: & $-$ & 0.3: & 0.3:\\
V42 & RRab     & $+27.0$ & $+21.6$ & 0.621752 & 18.9: & 18.4: & $-$  & 0.5: & $-$     & 0.5: & 0.5:\\
V43 & RRab     & $-16.9$ & $+22.6$ & 0.567657 & 18.5: & 18.1: & $-$  & 0.4: & $-$     & 1.3: & 1.1:\\
V44 & RRc      & $+19.1$ & $-48.0$ & 0.357335 & 18.6: & 18.3: & $-$  & 0.3: & $-$     & 0.5: & 0.5:\\
V45 & RRab     & $+5.0$  & $-77.4$ & 0.640064 & 18.6: & 18.2: & $-$  & 0.4: & $-$     & 0.5: & 0.4:\\
V46 & RRab$^a$     & $-47.7$ & $-14.9$ & 0.638804 & 18.5: & 18.0: & $-$  & 0.5: & $-$ & 0.2: & 0.2:\\
V47 & RRc      & $+47.8$ & $-12.5$ & 0.333933 & 18.1: & 17.8: & $-$  & 0.3: & $-$     & 0.6: & 0.5:\\
V48 & RRab$^a$     & $+27.8$ & $-19.1$ & 0.516412 & 18.2: & 17.9: & $-$  & 0.3: & $-$ & 0.5: & 0.4:\\
\end{tabular}
\end{minipage}
\end{table*}

We employed Layden's (1998) `light-curve templates' technique 
to assist us in deriving the RRab amplitudes. Layden only 
provided templates in the $V$ band. In order to derive the 
$B$ amplitudes, we utilized his $V$ templates as well, given the 
similarity in shape of the normalized light curves. We tested the 
accuracy of the procedure by applying the $V$ templates to a
sample of RRab stars with very well-defined $B$ light curves from 
the literature, encompassing a wide variety of light curve shapes. 
The stars included are V9 in 47~Tuc; V3, V7, V10 and V13 in M2 
(NGC~7089); V8 and V28 in M5; V4, V5 and V8 in M92 (NGC~6341); 
and the field star SW~Draconis. The data were obtained from 
the following sources. 47~Tuc: Carney, Storm, \& Williams (1993); 
M2: Lee \& Carney (1999); M5: Storm, Carney, \& Beck (1991); 
M92: Carney et al. (1992); SW~Dra: Jones et al. (1987). As a rule, 
we found that the $B$ amplitudes of these stars were successfully 
reproduced using the normalized $V$ templates from Layden. The 
mean difference between the values tabulated in the quoted 
papers and those derived using the $V$ template approach is 
essentially negligible, 
$\langle \Delta A_B\rangle = +0.002 \pm 0.017$~mag 
(standard deviation). However, we recommend that the $V$ templates 
are not used to infer mean magnitudes and colours, since there is 
a noticeable difference in shape between normalized $B$ and $V$ 
light curves at intermediate phases, $\phi \sim 0.5$. In Figs.~2 
and 3, we overplot the best-fitting light curves obtained from 
Layden's templates on data for the RRab stars with `regular' light 
curves (see below).   

The entries in Tables~\ref{Tab02},~\ref{Tab03} and~\ref{Tab04} have 
essentially the same meaning as in Table~\ref{Tab01}, though the 
following should be noted. In Tables~\ref{Tab02} and~\ref{Tab03}, 
new `V' numbers, starting from the last entry in Sawyer Hogg's 
(1973) catalogue, are assigned in order of discovery. For ease 
of reference, cross-identification with the numbers given in CFT91 
and BCSS97 is also provided. Not listed are the BCSS97 candidates 
1, 2, 4, 5, and 6; however, for future reference, we assign to these 
likely variable stars the numbers V25, V26, V28, V29, and V30, 
respectively. 

Specifically in regard 
to the new variable stars (V39--V48) that we have identified using 
{\sc isis} (Table~\ref{Tab04}), estimates of amplitudes inferred 
from the minimum and maximum values of the relative fluxes 
are given (see Sect.~2.3 for a discussion). For these stars, 
`mean' $B$ and $V$ magnitudes and 
$\langle B \rangle - \langle V \rangle$ colours are given for 
reference purposes only, and are based solely on our best $B,\,V$ 
image pair.  
All these uncertain values are marked with a colon in 
Table~\ref{Tab04} and must {\em not} be used except with the 
purpose of assessing the variability type. In any case, it is 
interesting to note that the RRab stars discovered using {\sc isis}
tend to have smaller amplitudes and longer periods than the 
bulk of the NGC~6229 RRab variables.

To check the above derived ephemerides, the intensity-mean 
$V$ magnitudes as a function of $(B-V)_{\rm mag}$ are overplotted 
on the CMD of NGC~6229 in Fig.~\ref{Fig05}. As pointed out by 
Preston (1961) and Sandage (1990), $(B-V)_{\rm mag}$ is the most 
appropriate colour to characterise the temperature of an `equivalent  
static star'. The NGC~6229 ridgelines are from Borissova et al. 
(1999); for the red HB, it represents the lower envelope of the 
distribution. 

Filled squares denote the RRc stars, and open circles 
correspond to RRab stars. The filled circles represent the known 
Population~II Cepheid in the cluster (V8), as well as V27 and 
V31---whose variability status we will discuss in more detail 
below. 

The mean $B$ and $V$ magnitudes for all RR Lyrae stars (except 
V32, which may be a field variable, and in any case is an outlier; 
see below) for which we have reliable {\sc daophot} light curves 
are $18.38\pm0.10$~mag and $18.03\pm0.08$~mag, respectively---where 
the standard deviation of the mean is given.  The latter 
value is in good agreement with the HB level 
($V_{\rm HB}=18.10\pm0.05$~mag) obtained by Borissova et al. (1999; 
see their Fig.~6) as an estimate of the lower envelope of the 
non-variable red HB stars' distribution in the NGC~6229 CMD. The 
mean colour of the RRc stars (except V32 and the ISM stars for which 
we do not have reliable {\sc daophot} photometry) is  
$\langle (B-V)_{\rm mag}\rangle_{\rm c} = 0.26\pm0.03$~mag, 
whereas 
$\langle (B-V)_{\rm mag}\rangle_{\rm ab} = 0.41\pm0.03$~mag 
for the RRab's.

Figure~\ref{Fig05} suggests that there is some overlap between 
the RRab's and non-variable red HB stars in NGC~6229. However, 
part of this effect appears to be due to the fact that the red 
HB `ridgeline' in Fig.~6 of Borissova et al. (1999) was indeed 
extended towards bluer colours than would be appropriate for a 
detailed comparison with RR Lyrae colours such as the one we 
are presently carrying out; a star-by-star analysis shows that 
the overlap is indeed relatively small. 
In order to determine the blue and 
red edges of the instability strip, we followed a procedure 
similar to that of Walker (1994). Accordingly, the blue edge of 
the instability strip comes from an average between the 
two bluest variables and the two reddest non-variables on the 
blue HB, giving $(B-V)_{\rm 0,\,BE} \simeq 0.22$~mag. The red 
edge is similarly derived by averaging over the five reddest 
variables and the five bluest among the non variables on the 
red HB, which gives $(B-V)_{\rm 0,\,RE} \simeq 0.45$~mag. 
Also, a `boundary' between the RRc and RRab variables can be 
placed at $(B-V)_{0} \simeq 0.32$~mag; there is little, if 
any, overlap between RRc's and RRab's in NGC~6229. A reddening 
value $E(B-V) = 0.01$~mag was assumed (Harris 1996). 

The periods of RRc stars are between $0\fd 264$ and $0\fd 398$, 
with a mean value $0\fd 32$. The periods of RRab stars are
between $0\fd 466$ and $0\fd 640$, with a mean value $0\fd 55$. 
In addition, the fraction of c-type RR Lyrae variables is 
$f_{\rm c} \simeq 0.24$. All these values are in excellent 
agreement with NGC~6229's classification as an Oosterhoff type 
I (OoI) GC, as first suggested by Mayer (1961; see also 
Castellani \& Quarta 1987).

\section{`Anomalous' stars, Blazhko variables and period changes}

RR Lyrae stars in GCs, particularly those pulsating in the fundamental 
mode (i.e., RRab's), often show non-repeating light curves, which
is characteristic of the Blazhko effect (e.g., Smith 1995). To 
identify possible Blazhko variables, we used the method derived by 
Jurcsik \& Kov\'acs (1996), which allows one to identify variables 
with `anomalous' light curves (often because of the Blazhko effect) 
from analysis of their Fourier decomposition parameters. In this 
method, a `deviation parameter' $D_m$ is introduced which measures 
the extent to which any given Fourier parameter `predicted' from 
empirical interrelations among the several other Fourier parameters 
(such relations being based on a large calibrating sample of RRab 
stars, mostly in the field) match the actual value of the Fourier 
parameter for the star. Jurcsik \& Kov\'acs argue that if the  
deviation in {\em any} of the Fourier parameters can be characterised 
by $D_m > 3$, the star has a `peculiar' light curve. More recently, 
Walker \& Kov\'acs (2000) have favoured a more restrictive criterion, 
$D_m > 5$. 

Fourier decompositions of the $V$ light curves were carried out 
following Simon \& Teays (1982). Fourier decomposition amplitudes 
$A_0$ to $A_{10}$ and phase differences $\phi_{21}$ to $\phi_{61}$ 
were explicitly calculated for use with the Kov\'acs \& Kanbur 
(1998) relations. This was done independently using both standard 
$V$ magnitudes (derived from {\sc daophot}) or ISM relative 
fluxes (derived from {\sc isis}; see Sect.~2). 
We excluded from our analysis stars 
which have prominent `gaps' in phase at the maximum of the light 
curve---cf. V10 and V39  in Figs.~3 and 4. We note 
that the method has been calibrated using light curves derived for 
standard $V$ magnitudes only (Jurcsik \& Kov\'acs 1996; Kov\'acs \& 
Kanbur 1998), so that the $D_m$ approach is not strictly applicable 
to identify `anomalous' stars using our ISM light curves. 
However, since the Fourier phases and the Fourier amplitudes 
(except $A_{0}$) do not appear to vary dramatically from one 
case to the other---in spite of some small differences---we shall 
assume that the calibrating equations remain valid in the case of 
{\sc isis}-based `$V$ magnitudes' as well. 

From this analysis we identified 15 stars with $D_{m} > 5$ which can 
accordingly be classified as variables with `peculiar' light curves: 
V4, V5, V6, V7, V18, V20, V33, V34, V35, V37, V38, V40, V41, V46 and 
V48. For some of these stars, as previously stated, light curves are 
available from Baade (1945) and Mannino (1960), thus allowing us to 
directly check for 
the possible presence of amplitude and/or period changes. To achieve 
this goal, Baade's photographic (ph) magnitudes were transformed to 
the Johnson $B$ filter using 15 comparison stars from Baade's Table~6,  
whereas Mannino's were placed by Mannino himself on Baade's system. 
The periods of these variables were re-derived following the same 
approach as for our CCD data (see Sect.~3 above). 

Comparison between the periods that we derived from Baade's (1945) 
and Mannino's (1960) data and those listed in Sawyer Hogg's (1973) 
catalogue reveals 
differences which are generally less than $0\fd 00009$. Since the 
estimated errors of our period determination are $\la 0\fd 00005$, 
we will consider these differences within the (combined) errors and 
thus consider the periods listed by Sawyer Hogg for these 15 stars 
as being satisfactory at $> 3\sigma$ (error of period determination). 
As previously stated, these periods were determined by Mannino  
and by Mayer (1961).

\begin{figure*}
 \centerline{\epsfig{figure=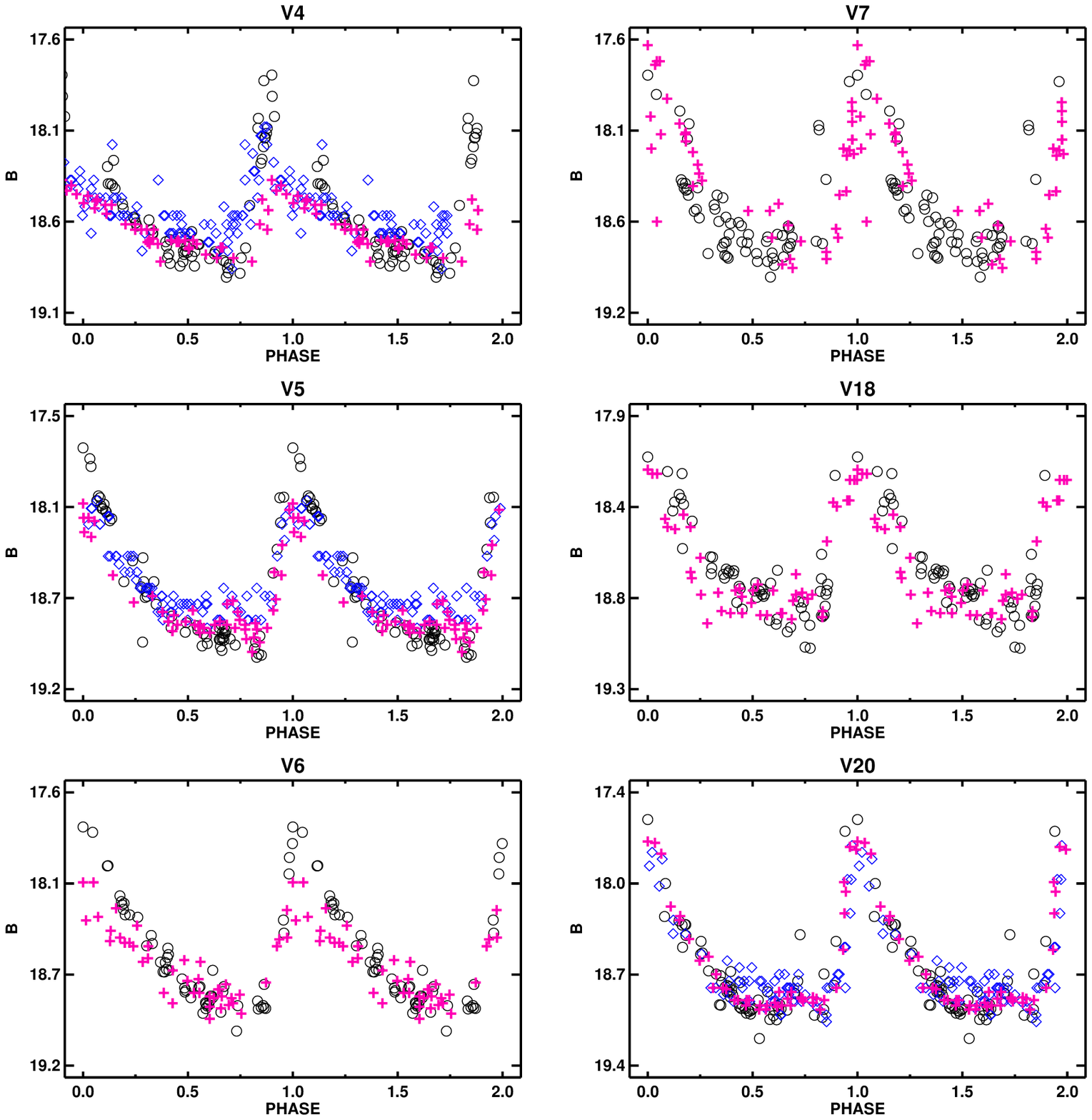}}
      \caption{$B$ light curves as a function of phase for Blazhko  
      variables. The plus signs and lozenges represent 
      Baade's (1945) and Mannino's (1960) photographic magnitudes,  
      respectively, transformed to the Johnson $B$ filter.}
      \label{Fig07}
\end{figure*}

Light curves for the `anomalous' variables V4, V5, V6, V7, V18 
and V20 are shown in Fig.~\ref{Fig07}. Baade's (1945) data are 
overplotted with plus signs, and Mannino's data with lozenges. 
Except for V7---no data available from Mannino (1960)---the 
amplitude changes for all these stars are obvious; they are 
clearly Blazhko variables. In Table~\ref{Tab05}, we quantify the 
change in $B$ amplitudes as derived from Baade's and Mannino's 
photographic data and our CCD material. The error range of 
amplitude values obtained as a combination of photometric errors, 
period determination and standard fit of the light curves is not 
larger than $0.05-0.07$~mag; hence the $B$ amplitude changes appear 
to be significant at $> 3\sigma$. Interestingly, all amplitudes are 
found to be increasing. 

For classification purposes, in Tables~1--4 we have assumed that 
all RRab variables with $D_m > 5$ are Blazhko variables (cf. 
Jurcsik \& Kov\'acs 1996, especially their Sect.~4.2). However, 
the possibility cannot be excluded that there are Blazhko stars 
whose $D_m$ values never become larger than $\simeq 3$, especially 
if the amplitude modulation is not too strong (Jurcsik 2000). In 
fact, we note that the RRab Lyrae V34 in M3 {\em is} a Blazhko 
variable (Carretta et al. 1998; Clement 1997 and references 
therein), whereas Kaluzny et al. (1998) have recently obtained  
$D_m = 1.9$ from their light curve for this star. Conversely, it 
is well known that there {\em are} stars with $D_m > 3$ which 
are {\em not} Blazhko variables (see Table~8 in Jurcsik \& 
Kov\'acs 1996).

\begin{table*}
 \begin{minipage}{160mm}
\caption {$B$ amplitude differences of Blazhko variables}
\begin{tabular} {lccccc}

 Name& $A_{B}$      & $A_{B}$        & $A_{B}$     & $\Delta A_{B}$     & $\Delta A_{B}$       \\
     & (Baade 1945) & (Mannino 1960) & (this work) & (Baade--this work) & (Mannino--this work) \\ 
        V4  &  0.53    &  0.70 & 1.15    &     $-0.62$  & $-0.45$ \\
        V5  &  0.70    &  0.79 & 1.30    &     $-0.60$  & $-0.51$ \\ 
        V6  &  0.68    &  $-$  & 1.15    &     $-0.47$  & $-$     \\
        V18 &  0.66    &  $-$  & 0.90    &     $-0.24$  & $-$     \\
        V20 &  1.14    &  1.14 & 1.60    &     $-0.46$  & $-0.46$ \\

\end{tabular}
\end{minipage}
\label{Tab05}
\end{table*}

In general, comparison between the periods listed in Sawyer Hogg's 
(1973) catalogue and those derived from our CCD data shows that 
most of them agree to within $0\fd 00009$. Possible significant 
period changes are found for only two variables: V7 (RRab) and V17 
(RRc). For V7 (V17), Sawyer Hogg lists a period of $P = 0\fd 506980$
($0\fd 324880$). Using Baade's (1945) data we obtain 
$P = 0\fd 506978$ ($0\fd 324872$), whereas the period obtained from 
our CCD data is $P = 0\fd 501600$ ($0\fd 325193$). No significant 
improvement is found for V17 using Mannino's (1960) data, due to 
its poor quality---Mannino himself was unable to find 
a period for this star.
The calculated difference $\Delta\,P_{\rm ph-CCD}$ for V7 and V17 
are $0\fd 00538$ and $-0\fd 000313$, respectively. Unfortunately, 
the reality of these period changes cannot be considered definite, 
since we do not have enough data to construct $O-C$ residuals and 
determine the type and rate of period changes. It should be of 
interest to assemble all the datasets available for this cluster 
(cf. Davis 1917; Baade 1945; Sawyer 1953; Mannino 1960; and the 
present work) in order to study the period change rates in 
NGC~6229, given the constraints that these can provide on stellar 
evolution, particularly on the HB phase (Smith 1995 and references 
therein). This will be the subject of a future paper.

\section{The Bailey diagram revisited}

\begin{figure*}
 \vspace{0.2in}
 \centerline{\epsfig{figure=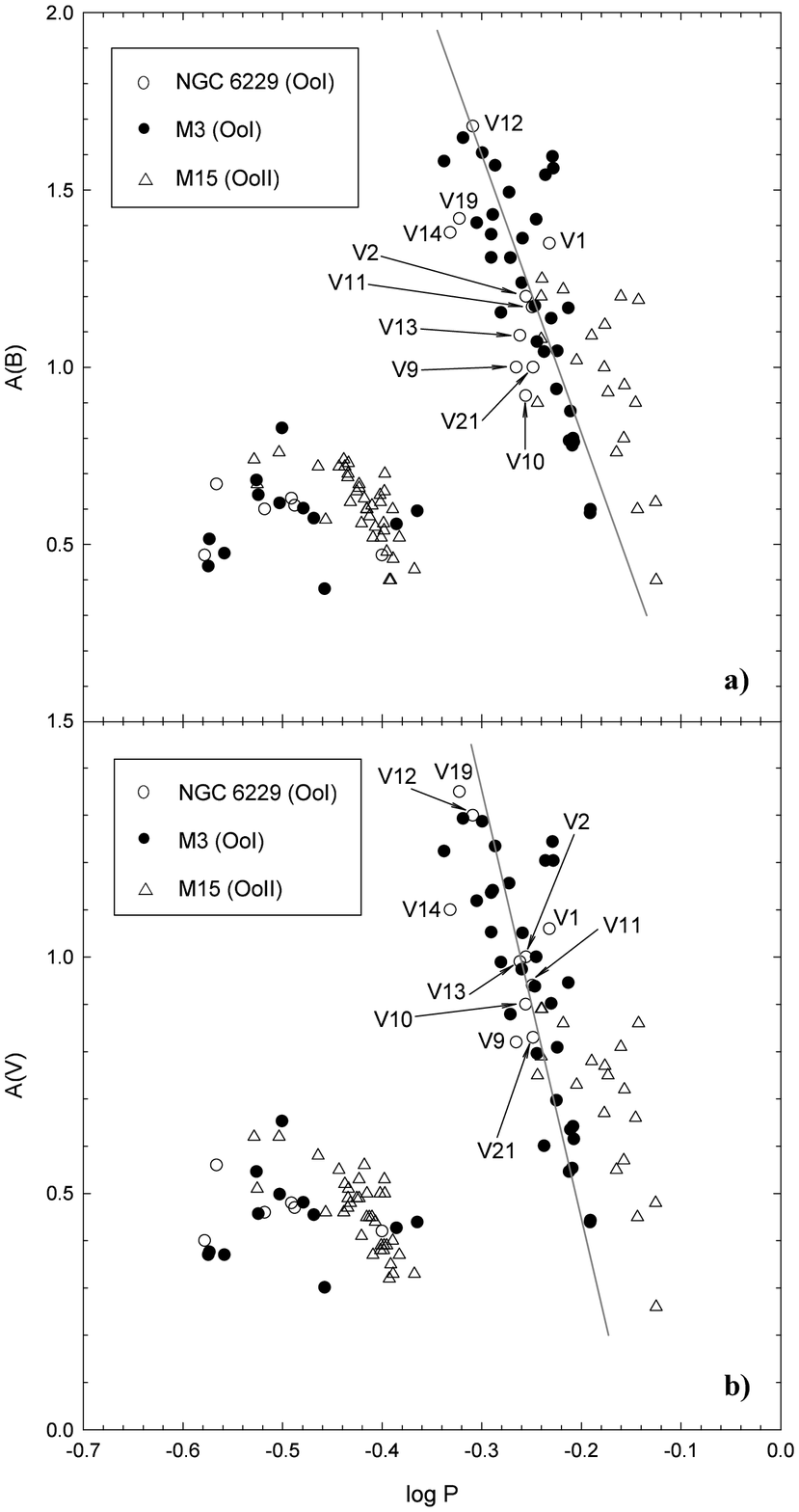, height=6.75in, width=4in}}
      \caption{Bailey diagrams for the RR Lyrae stars  
      in NGC~6229 (open circles). Panel~a) shows the 
      period-$B$ amplitude diagram, while panel~b) shows the 
      period-$V$ amplitude distribution. For comparison purposes, 
      data for the OoI GC M3 (filled circles) and the OoII GC M15 
      (filled triangles) are also displayed. Blazhko and/or 
      variables with large $D_m$ values (Jurcsik \& Kov\'acs 
      1996) were avoided (see text). Individual RRab stars in 
      NGC~6229 are indicated by their names. The solid line in
      each plot corresponds to the `OLS bisector' best fit to the
      M3 RRab data (see text). 
      }
   \label{Fig08}
\end{figure*}

In Fig.~\ref{Fig08}, the Bailey diagram is plotted. Individual 
stars are identified for ease of reference. 
Panel~a) shows the period-$B$ amplitude distribution, whereas 
panel~b) corresponds to the period-$V$ amplitude distribution. 
RRc variables (short periods and low amplitudes) stand out 
clearly from their RRab counterparts in this diagram.
NGC~6229 variables from Tables~1, 2, and 3 are shown 
as open circles. Only RRab stars with $D_m < 5$ are displayed;
the only exception is V10, for which we have not computed a 
$D_m$ value (cf. Sect.~4). Comparison between our light curves 
and those based on the data from Baade (1945) and Mannino 
(1960) does not show any sign of amplitude variations at the 
$\approx 0.1$~mag level, which suggests that V10 is not a 
Blazhko variable.   

Also plotted are M3 stars (data from Carretta et al. 1998) 
and M15 (NGC~7078) variables (data from Bingham et al. 1984, as 
selected by Catelan 1998). As well known, these GCs are typical 
representatives of the OoI and OoII classes, respectively. 
M3 stars are shown as solid circles, whereas open triangles 
indicate the M15 variables. As was done in the case of M15 by 
Catelan, we have removed all M3 stars that 
Carretta et al. explicitly note to be (possible) Blazhko 
variables, as well as M3 variables identified as showing the 
Blazhko effect in Clement's (1997) catalogue. In addition, we 
cross-checked 
the Carretta et al. and Kaluzny et al. (1998) lists to identify 
possible stars in common between the two studies with large 
$D_m$ values. None was found. As already mentioned though, we 
found that V34, which {\em is} a Blazhko variable according to 
Carretta et al. and Clement's catalogue, has a very small $D_m$ 
value---which might have led one to conclude that it has a 
`normal' light curve. V34 was also excluded from Fig.~\ref{Fig08}.  

It is immediately clear that there is a period shift (at fixed 
amplitude) between NGC~6229 and M3 RRab variables, in the 
sense that M3 stars have longer periods than their NGC~6229 
counterparts. This is more easily noticed when we overplot the 
lines that best fit the M3 data, obtained following the `OLS 
bisector' method as appropriate in this case (cf. Isobe et al. 
1990): 

\begin{displaymath}
 \log\,P_{\rm ab}({\rm M3}) = -0.097 - 0.127\,A_B, 
\end{displaymath}  

\begin{displaymath}
 \log\,P_{\rm ab}({\rm M3}) = -0.110 - 0.151\,A_V. 
\end{displaymath}  

\noindent Most NGC~6229 RRab stars with regular light curves 
fall below these best-fitting M3 lines, indicating shorter RRab
periods at a given amplitude than is the case for M3. 
Surprisingly, the size of the resulting period shifts 
is not the same when the $B$ and $V$ amplitudes are considered. 
In the former case, one finds a mean period shift for the 10 
RRab stars (in the sense NGC~6229 minus M3) of 
$\langle \Delta\,\log\,P(A_B)\rangle = -0.021 \pm 0.009$ 
(standard error of the mean); in the latter, one has instead
$\langle \Delta\,\log\,P(A_V)\rangle = -0.008 \pm 0.008$. 
Given the slopes of the M3 period-amplitude relations, such a 
difference could be accounted for if the $V$ amplitudes were 
too large by $\approx 0.09$~mag, or if the $B$ amplitudes were 
too small by $\approx 0.11$~mag (or some combination thereof). 
We believe that better data will be necessary before the cause 
of this difference can be fully explained. We shall assume that 
the mean between the two values, $-0.015$, is a good indicator 
of the actual period shift between the two clusters.  

Fundamental-mode blue amplitudes and 
temperatures appear to be closely related (Sandage 1981a, 1981b; 
Jones et al. 1992; Catelan 1998; Sandquist 2000). Therefore, 
period shifts at constant blue amplitude can be interpreted as 
period shifts at (approximately) constant temperature. From 
the period-mean density relation (van Albada \& Baker 1971), 
it thus follows that the detected 
period shift must be due to luminosity and/or mass differences 
between M3 and NGC~6229 variables. Assuming similar masses for 
the variables in the two clusters, this implies that the M3 
variables might be brighter than their NGC~6229 counterparts 
by $\approx 0.05$~mag. It is worth noting that such luminosity 
differences at a fixed metallicity, if indeed present, can only 
be caused by variations in `second-parameter' candidates other 
than age or mass loss on the red giant branch, with possible 
implications for our understanding of the `HB bimodality'
phenomenon (e.g., Catelan, Sweigart, \& Borissova 1998). 
On the other hand, the period shift could also, at least in 
principle, be due to a mass difference between the RR Lyrae 
variables in the two GCs. At a fixed luminosity, a mass difference 
$\Delta\,M \approx 0.03\,M_{\odot}$, in the sense that M3 variables 
are less massive, would be required in this case. 

We recall, from Borissova et al. (1999), that NGC~6229 appears 
to have virtually the same metallicity as M5. On the 
other hand, it is quite well established that M5 is more 
metal-rich than M3 by $\simeq 0.2-0.3$~dex (see Jurcsik 1995 
and references therein). Therefore, contrary to Clement \& 
Shelton (1999), our results suggest that there is no universal, 
metallicity-independent `OoI line' in the period-amplitude 
diagram. However, we cannot exclude some systematic effect 
due to our relatively small number of RRab stars with regular 
light curves and/or to uncertainties in the derived amplitudes. 
Better data for NGC~6229 are needed before 
this issue can be settled with certainty. We mention, in 
passing, that Kaluzny et al. (2000) have recently claimed 
that the period-$V$ amplitude diagram is indeed a function 
of metallicity, based on their M5 analysis.

\section{Physical properties of RR Lyrae stars from their
         Fourier parameters}

\subsection{RRc stars}

Simon \& Clement (1993) used light curves of RRc stars obtained from 
hydrodynamic pulsation models to derive equations to calculate mass, 
luminosity, temperature and a `helium parameter' as a function of the 
Fourier phase difference $\phi_{31}$ and period. 
From their equations~(2), (3), 
(6) and (7), we computed these quantities using the Fourier 
decomposition parameters obtained from our CCD light curves. They are 
given in Table~\ref{Tab06}. Note that the `helium parameter' $y$ does 
{\em not} provide a good description of the helium abundance (Simon 
\& Clement 1993).

\begin{table}\tabcolsep=3pt\small
\caption {Physical parameters for RRc variables in NGC~6229}
\begin{tabular} {lcccccccc}

Name & $A_{0}$  &$A_{1}$ & $\phi_{31}$ & Mass & $\log\,(L/L_{\odot})$ & $T_{\rm eff}$ & $y$ & $\sigma(\phi_{31})$ \\
     &          &        &             & ($M_{\odot}$) &  & (K)           &     &  \\ 

V15  &  18.14 & 0.249 & 3.512  & 0.512 & 1.62 & 7526  & 0.30 & 0.04\\
V16  &  18.02 & 0.178 & 3.077  & 0.625 & 1.72 & 7297  & 0.26 & 0.26\\
V17  &  17.96 & 0.212 & 3.120  & 0.620 & 1.72 & 7294  & 0.26 & 0.07\\
V23  &  17.93 & 0.175 & 4.146  & 0.532 & 1.75 & 7178  & 0.26 & 0.03\\    
V24  &  18.08 & 0.206 & 3.239  & 0.581 & 1.68 & 7379  & 0.28 & 0.04\\
V32  &  18.21 & 0.076 & 3.513  & 0.602 & 1.76 & 7188  & 0.25 & 0.17\\ 

\end{tabular}
\label{Tab06}
\end{table}

\begin{table}\tabcolsep=3pt\small
\caption {Physical parameters for RRc variables from ISM light curves}
\begin{tabular} {lccccccc}

Name &$A_{1}$ & $\phi_{31}$ & Mass & $\log\,(L/L_{\odot})$ & $T_{\rm eff}$ & $y$ & $\sigma(\phi_{31})$ \\
     &        &             & ($M_{\odot}$) &  & (K)           &     &  \\ 

V15  &   0.184 & 3.499  & 0.514   & 1.62    & 7524  & 0.30 & 0.03\\
V16  &   0.140 & 3.499  & 0.562   & 1.70    & 7337  & 0.28 & 0.04\\
V17  &   0.175 & 3.564  & 0.555   & 1.70    & 7335  & 0.28 & 0.02\\
V23  &   0.192 & 3.512  & 0.624   & 1.79    & 7119  & 0.25 & 0.04\\ 
V24  &   0.158 & 3.832  & 0.500   & 1.65    & 7435  & 0.30 & 0.03\\
V32  &   0.076 & 3.320  & 0.637   & 1.78    & 7158  & 0.25 & 0.08\\ 
V36  &   0.152 & 3.649  & 0.487   & 1.60    & 7569  & 0.31 & 0.04\\
V44  &   0.168 & 3.171  & 0.644   & 1.76    & 7199  & 0.26 & 0.03\\
V47  &   0.205 & 3.611  & 0.555   & 1.70    & 7312  & 0.28 & 0.03\\

\end{tabular}
\label{Tab07}
\end{table}

We exclude from our analysis V36 due to its irregular $V$ light curve and 
hence large $\phi_{31}$ error.
V16 is another variable with an irregular $V$ light curve and relatively  
large $\sigma(\phi_{31})$.

Excluding V32 (see below), the unweighted mean values and 
standard deviations of the mass $M$, 
luminosity $\log\,(L/L_{\odot})$, effective temperature $T_{\rm eff}$ and 
`helium parameter' $y$ are: $0.57\pm0.05\,M_{\odot}$, $1.70\pm0.05$, 
$7335\pm129$~K and 
$0.27\pm0.02$, respectively. For comparison purposes, 
we find, from the latest VandenBerg et al. (2000) $\alpha$-enhanced 
zero-age HB (ZAHB) models, at $T_{\rm eff} = 7300$~K, 
the following values: for ${\rm [Fe/H]} \simeq -1.4$ (NGC~6229 
metallicity in the Zinn \& West 1984 scale; cf. Borissova et al. 1999), 
$M_{\rm ZAHB} \simeq 0.62\,M_{\odot}$, 
$\log\,(L/L_{\odot})_{\rm ZAHB} \simeq 1.62$; 
for ${\rm [Fe/H]} \simeq -1.1$ (NGC~6229 
metallicity in the Carretta \& Gratton 1997 scale; cf. Borissova et al. 
1999), $M_{\rm ZAHB} \simeq 0.59\,M_{\odot}$, 
$\log\,(L/L_{\odot})_{\rm ZAHB} \simeq 1.61$. As can be seen, the Simon 
\& Clement (1993) method favours the `long' Pop.~II distance scale, 
providing higher HB luminosities than those predicted by VandenBerg 
et al. (2000).  

The same analysis was carried out using the ISM ({\sc isis})
light curves, and the 
results are presented in Table~\ref{Tab07}. Note the remarkable decrease 
in $\sigma(\phi_{31})$ for V16 and V32 from Table~\ref{Tab06} to 
Table~\ref{Tab07}. We were 
also able to successfully analyse V36 with the ISM light curves, 
obtaining a relatively small $\sigma(\phi_{31})$. The unweighted mean 
values and standard deviations 
of the mass $M$, luminosity $\log\,(L/L_{\odot})$, effective 
temperature $T_{\rm eff}$ and `helium parameter' $y$ are: 
$0.56\pm0.05\,M_{\odot}$, $1.69\pm0.06$, $7332\pm157$~K and 
$0.28\pm0.02$, respectively. 
As can be seen, the {\sc isis} values are in good agreement 
with those obtained using the standard $V$ light curves. 

How does this compare with M3? For this GC, Kaluzny et al. (1998) 
have found the following mean values: 
$\langle M\rangle = 0.59\,M_{\odot}$;  
$\langle \log\,(L/L_{\odot})\rangle = 1.71$;
$\langle T_{\rm eff} \rangle = 7315$~K. This suggests masses lower 
in NGC~6229 than in M3 by $\approx 0.02-0.03\,M_{\odot}$ (in the 
{\em opposite} sense to that required to account for the detected 
period shift between the two clusters), and luminosities lower 
in NGC~6229 than in M3 by $0.03-0.05$~mag (as needed to account for 
the period shift between these clusters). One should note, however, 
that this method may need further testing and calibration, given 
the empirical results recently presented by Kov\'acs (1998).

\begin{table}\tabcolsep=3pt\small
\caption {Physical parameters for RRab variables in NGC~6229}
\begin{tabular} {lcccccccc}

Name&$A_{0}$ &$A_{1}$ & $\phi_{31}$ & $\phi_{41}$&[Fe/H]&$M_{V}$&$T_{\rm eff}$&$\sigma({\rm [Fe/H]})$\\
    &        &        &             &            &      &       & (K)         &  \\

V1  &  17.934 &  0.400 &  4.949 &  1.182 & $-1.54$ & 0.72 & 6438  & 0.09\\
V2  &  18.078 &  0.371 &  4.800 &  1.189 & $-1.58$ & 0.76 & 6431  & 0.08\\
V9  &  18.029 &  0.273 &  5.037 &  1.119 & $-1.19$ & 0.85 & 6493  & 0.15\\
V11 &  18.181 &  0.288 &  4.994 &  1.279 & $-1.27$ & 0.81 & 6478  & 0.09\\
V12 &  18.043 &  0.369 &  4.483 &  4.591 & $-1.45$ & 0.78 & 6444  & 0.10\\
V13 &  18.106 &  0.293 &  4.858 &  1.373 & $-1.46$ & 0.82 & 6410  & 0.07\\
V14 &  17.926 &  0.360 &  4.688 &  0.787 & $-1.25$ & 0.88 & 6595  & 0.09\\
V19 &  18.067 &  0.452 &  4.883 &  1.005 & $-1.04$ & 0.84 & 6654  & 0.08\\
V21 &  18.025 &  0.248 &  5.392 &  1.328 & $-0.83$ & 0.87 & 6355  & 0.11\\

\end{tabular}\label{Tab08}
\end{table}

\begin{table}\tabcolsep=3pt\small
\caption {Physical parameters for RRab variables in NGC~6229 from the ISM light curves}
\begin{tabular} {lcccccccc}

Name&[Fe/H]& $M_{V}$&$T_{\rm eff}$&$\sigma$([Fe/H])\\

V1  & $-1.65$ & 0.73 & 6370  & 0.09\\
V2  & $-1.61$ & 0.77 & 6423  & 0.08\\
V5  & $-1.53$ & 0.81 & 6462  & 0.13\\
V6  & $-1.41$ & 0.79 & 6308  & 0.12\\
V7  & $-1.65$ & 0.83 & 6380  & 0.16\\
V9  & $-1.46$ & 0.81 & 6376  & 0.15\\
V11 & $-1.40$ & 0.83 & 6437  & 0.09\\
V12 & $-1.44$ & 0.79 & 6431  & 0.10\\
V13 & $-1.29$ & 0.83 & 6427  & 0.07\\
V14 & $-1.27$ & 0.89 & 6525  & 0.09\\
V19 & $-1.68$ & 0.82 & 6446  & 0.08\\
V20 & $-1.54$ & 0.81 & 6553  & 0.19\\
V21 & $-0.93$ & 0.87 & 6372  & 0.11\\
V33 & $-1.33$ & 0.85 & 6025  & 0.09\\
V34 & $-1.48$ & 0.81 & 6356  & 0.13\\
V38 & $-1.45$ & 0.84 & 6487  & 0.18\\
V40 & $-0.81$ & 0.84 & 6227  & 0.15\\
V42 & $-1.75$ & 0.81 & 6379  & 0.17\\
V43 & $-1.74$ & 0.74 & 6344  & 0.13\\
V45 & $-0.85$ & 0.85 & 6322  & 0.12\\

\end{tabular}\label{Tab09}
\end{table}

\subsection{RRab stars}

Jurcsik \& Kov\'acs (1996), Kov\'acs \& Jurcsik (1996, 1997) and 
Kov\'acs 
\& Kanbur (1998) obtained empirical formulae relating the stellar 
metallicities, absolute magnitudes and temperatures to Fourier 
decomposition parameters for RRab stars with `regular' light curves.  
The only model-dependent ingredients in their calibrations are the 
zero point of the HB luminosity scale (adopted from Baade-Wesselink 
studies) and the colour-temperature transformations (obtained from 
static model atmospheres). 

The physical parameters of NGC~6229 RRab Lyrae variables obtained 
from this method are given in Table~\ref{Tab08}. The unweighted mean 
value (and corresponding standard deviation)  
of [Fe/H] derived from $\phi_{31}$ is $-1.26\pm0.24$~dex. Note that this 
value is in the Jurcsik (1995) scale. 
Likewise, the mean absolute magnitudes 
and temperatures are $\langle M_{V}\rangle=0.82\pm0.05$~mag and 
$\langle T_{\rm eff}\rangle=6478\pm93$~K. Again, similar values 
are obtained from the ISM light curves: 
$\langle {\rm [Fe/H]\rangle} = -1.41\pm0.28$~dex, 
$\langle M_{V}\rangle = 0.81\pm0.04$~mag and 
$\langle T_{\rm eff}\rangle=6383\pm116$~K. 
The faint HB level is a reflection of the adoption of the Baade-Wesselink 
luminosity zero point in the calibration of this method (see Jurcsik \& 
Kov\'acs 1999 for a recent discussion). 

Comparison with the mean absolute magnitude of the RRab Lyrae in 
M3 recently obtained by Kaluzny et al. (1998) using the same approach, 
$\langle M_{V}\rangle=0.78\pm0.01$~mag, is again suggestive of a 
slightly fainter HB level in NGC~6229 than in M3, as required in order 
to account for a period shift between the RRab variables in the two   
clusters. For an analysis of possible problems related to this method, 
the reader is refered to Koll\'ath, Buchler, \& Feuchtinger (2000) and 
Jurcsik \& Kov\'acs (1999).

\section{Comments on individual stars}

\begin{figure}
 \centerline{\epsfig{figure=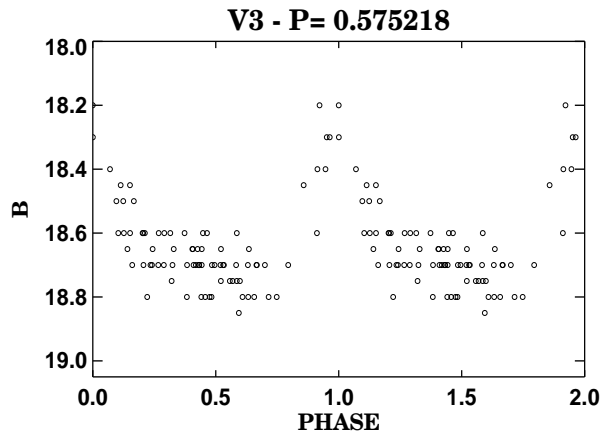, height=2.35in, width=8.5cm}}
      \caption{$B$ light curve for V3, obtained from Mannino's (1960) 
      ph data transformed to the Johnson system using the indicated 
      period. 
      }
      \label{Fig14}
\end{figure}

\noindent {\bf V3}: Although data (of low quality) have been provided 
for this star by Baade (1945) and by Mannino (1960), no period has been 
published to date. We note that our CCD investigation 
does not provide help in resolving the situation, since V3 is not in 
our observed field. However, we have investigated Mannino's data and 
found that it was possible to fold his observations with a period 
$P=0\fd 575218$. Fig.~\ref{Fig14} shows the first light curve for 
this star, which is clearly an RRab Lyrae variable. Due to its 
distance from the cluster centre and poor quality of the light curve, 
V3 was not considered when deriving mean cluster properties.

\noindent {\bf V12}: CFT91 noted that this RRab star, in comparison with 
the other NGC~6229 variables, `seems to be much too bright and too blue'. 
They suspected that this could be due to crowding, hence blending. 

Careful inspection of our images does reveal that V12 has a very close 
`companion'. This is the main reason why the ISM light curves for this 
star (Figs.~\ref{Fig03} and \ref{Fig04}) are significantly better than 
their {\sc daophot} counterparts (Figs.~\ref{Fig01} and \ref{Fig02}): 
{\sc isis} encountered no difficulty at removing this close 
companion. Note that, unlike in CFT91, the mean magnitudes and colours 
for this star (Table~1) do not deviate substantially from the remainder 
of the sample.

\subsection{RR Lyrae variables}

\noindent {\bf V32}: This RRc star is the faintest overtone variable in 
the NGC~6229 CMD (Fig.~\ref{Fig05}). V32, at 
$\langle V\rangle = 18.27$~mag, is almost 0.2~mag fainter than the lower 
envelope of the red HB of the cluster. 
On the other hand, the Fourier decomposition analysis suggests 
that V32 is one of the two {\em intrinsically brightest} RRc 
variables in the NGC~6229 field (cf. Tables~6 and 7). In spite of V32's
proximity to the cluster centre (Fig.~\ref{Fig06}), these properties 
suggest that V32 may be a non-member. Accordingly, all average 
values reported in this paper do not take V32 into account.

\begin{figure}
 \centerline{\epsfig{figure=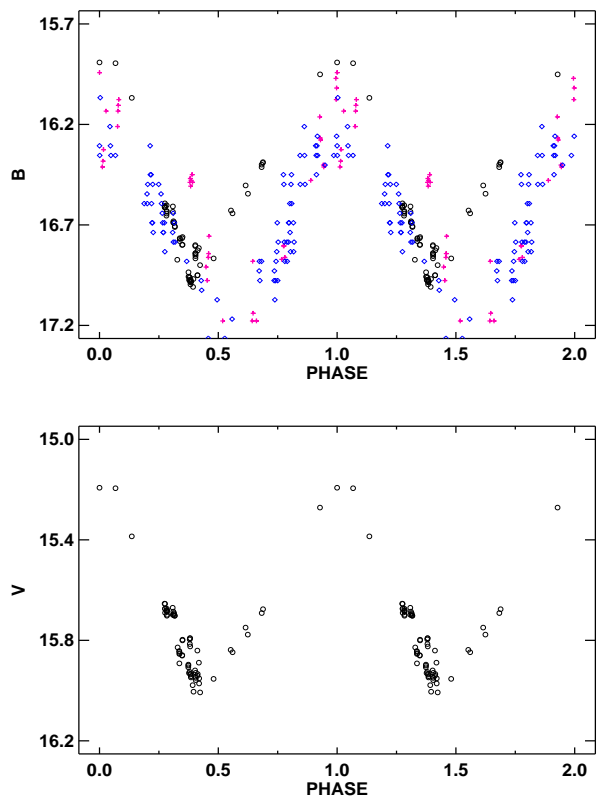, height=4.35in, width=8.5cm}}
      \caption{Light curves as a function of phase for V8. The plus  
      signs and lozenges represent Baade's (1945) and Mannino's (1960) 
      photographic magnitudes, respectively, transformed to the Johnson 
      $B$ filter as described in the text.
      }
      \label{Fig10}
\end{figure}

\subsection{Population II Cepheids} 

\noindent {\bf V8}: As previously stated, Davis (1917) found V8 to be 
one of the brightest stars in NGC~6229, and Baade (1945) classified 
it as a `long-period Cepheid'. Mannino (1960) obtained a period 
$P=14\fd 845093$, thus assigning to it W~Virginis status. The period 
we estimate from our new CCD data is slightly shorter: $P=14\fd 840457$. 
The period, $B$ amplitude and light curves (Fig.~\ref{Fig10}) are all
consistent with its classification as a W~Vir variable. 
In comparison with the typical colours of W~Vir stars in GCs, V8,  
with $(B-V)_{\rm 0, mag} \simeq 0.77$~mag, seems to be among the 
reddest: from Harris (1985), one might have expected a colour 
$(B-V)_0 \simeq 0.52$~mag given this star's period, whereas from Nemec, 
Nemec, \& Lutz (1994)  a colour $(B-V)_0 \simeq 0.6$~mag would seem 
more appropriate. Using the distance modulus from Harris (1996) and 
given that $\langle V\rangle = 15.53$~mag, its absolute magnitude is 
found to be $\langle M_V\rangle \simeq -1.85$~mag, which places it among 
the fundamental Pop.~II variables in the Nemec et al. classification 
scheme.

\begin{figure}
 \centerline{\epsfig{figure=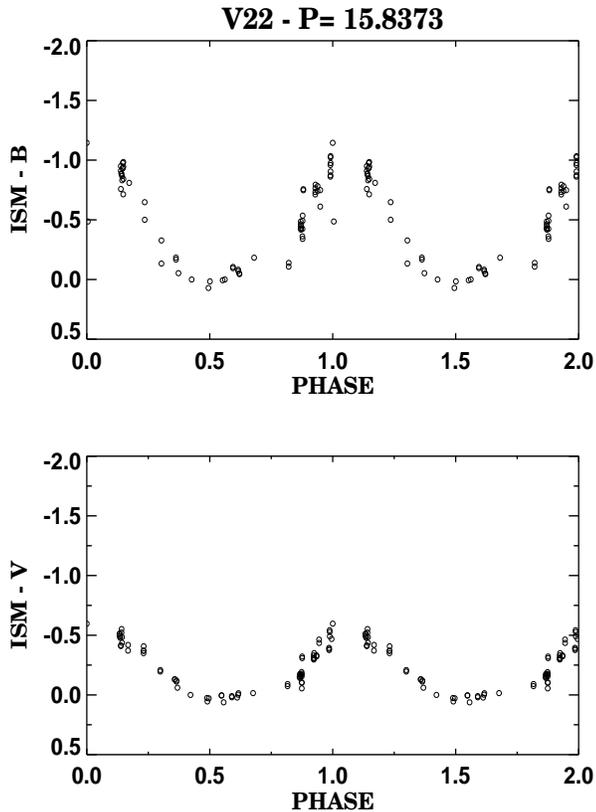, height=4.35in, width=8.5cm}}
      \caption{$B$, $V$ light curves for V22, obtained using the  
      ISM method ({\sc isis}) as applied to our shortest-exposure 
      frames.  
      }
      \label{Fig11}
\end{figure}

\noindent {\bf V22}: No light curve has ever been published for V22, 
which is located very close to the core of the cluster according to 
the coordinates provided by Sawyer (1953). To the best of our 
knowledge, no follow-up observations were carried out for this star 
since Sawyer reported on its discovery, although Sawyer Hogg's (1953) 
catalogue mentions that this variable is `probably slow' (i.e., 
has a long period). 

In order to assess the variability type and attempt to determine a
period for this star, we have used exclusively our 10 best and 
shortest (exposure times 60~sec) ISM-subtracted frames, since we 
expected that any star that could be found by Sawyer (1953) at the 
very core of the cluster using her photographic material had to be 
very bright, and hence probably only seen by Sawyer on her frames 
obtained with the shortest exposures. This expectation was also 
reinforced by Sawyer's remark that the star's `magnitude indicates 
that its period may be greater than a day'. 

The ISM method revealed {\em no} variable star at the position 
reported by Sawyer Hogg (1973). However, relatively close to her
coordinates, at $X = +3.8\arcsec$ and $Y = -10.5\arcsec$, a very 
bright variable star was found. We suggest that this is probably 
a rediscovery of Sawyer's (1953) V22, and that the previous 
coordinates for this star were incorrect. 

Since we also have $\approx 10$ {\sc daophot} magnitudes measured 
on our short (60~sec) frames, we attempted to derive a transformation 
between ISM relative fluxes and standard $B$, $V$ magnitudes for 
this star. We found that it was necessary to multiply the V22 ISM 
fluxes by 0.00002 and 0.0000046 in order to obtain magnitudes in 
$B$ and $V$, respectively. Note that these coefficients are 
different from those obtained for the RR Lyrae variables in 
Section~2.3. One possible reason for such a difference is related 
to the sky subtraction in this extremely crowded region of the 
cluster (see Fig.~1). Using $B$, $V$ magnitudes thus derived, 
light curves are shown in Fig.~\ref{Fig10}. As can be seen, a 
period $P=15\fd 8373$ gives an acceptable light curve for the 
star, suggesting that it is a W~Vir variable.  

The derived photometric parameters for this star are very uncertain, 
but they allow us to constrain the variability status of V22. The 
colour of the star is $(B-V)_0 \sim 0.9$~mag, making it the reddest 
among all known NGC~6229 variables. From Harris (1985), one would 
expect a colour $(B-V)_0 \simeq 0.52$~mag given this star's period, 
assuming it to be a W~Vir variable, whereas from Nemec et al. 
(1994) a colour $(B-V)_0 \simeq 0.6$~mag would seem more appropriate. 
Using the distance 
modulus from Harris (1996), V22's absolute magnitude is found to be 
$\langle M_V\rangle \sim -2.3$~mag. This would place V22 among the 
overtone Pop.~II Cepheids, following the Nemec et al. classification 
scheme. Note, however, that McNamara (1995) has questioned the 
reality of the subdivision of Pop.~II Cepheids into fundamental and 
overtone pulsators by Nemec et al. (1994). 

In summary, we confirm that V8 and V22 are most likely W~Vir 
variables, but more data are needed, especially in the case of V22,
to verify their colours.

\subsection{Other variable stars}

Two additional stars, V27 and V31, have average magnitudes placing them 
above the HB level of NGC~6229 (see Fig.~\ref{Fig05}). 
These bright variables were both discovered by BCSS97.

\begin{figure}
 \centerline{\epsfig{figure=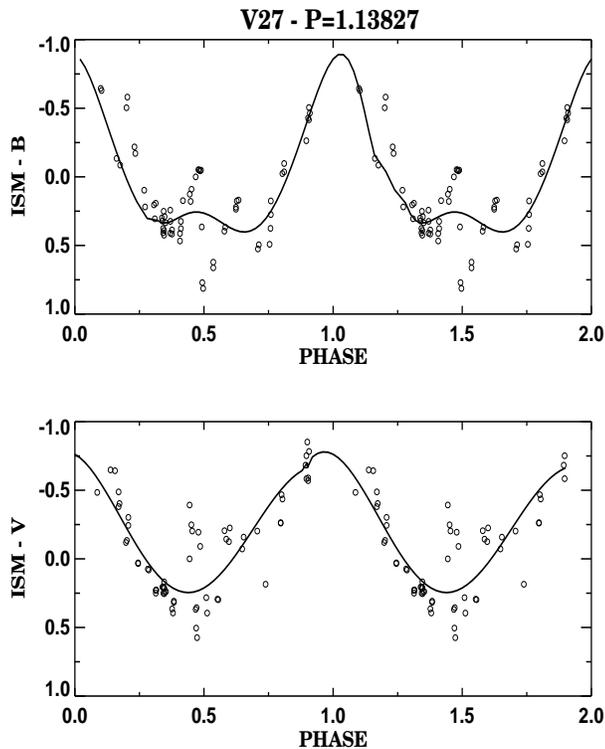, height=4.0in, width=8.5cm}}
      \caption{ISM light curves for V27, obtained using {\sc isis}.}
      \label{Fig12}
\end{figure}

\noindent {\bf V27}: V27 has a period $P=1\fd 13827$. The light curves 
of V27 in the relative ISM $B,\,V$ fluxes are presented 
in Fig.~\ref{Fig12}. V27 has period, colour and luminosity appropriate 
for its classification as a Pop.~II Cepheid of the BL Herculis type. 
However, the light curve is not well defined, and we regard the derived 
period uncertain, as well as its BL~Her classification.

\noindent {\bf V31}: From the analysis of {\sc daophot} magnitudes, V31 
has a period $P = 0\fd 715208$. A different period is derived on the basis 
of the ISM fluxes; in this case, we find $P=0\fd 698893$. V31 appears to 
lie on an extension of the BL~Her period-colour relation (Nemec et 
al. 1994) towards bluer colours, and its luminosity would be consistent 
with this interpretation. V31 might also be classified as an `AHB1' star 
(Diethelm 1983). 

\begin{figure}
 \centerline{\epsfig{figure=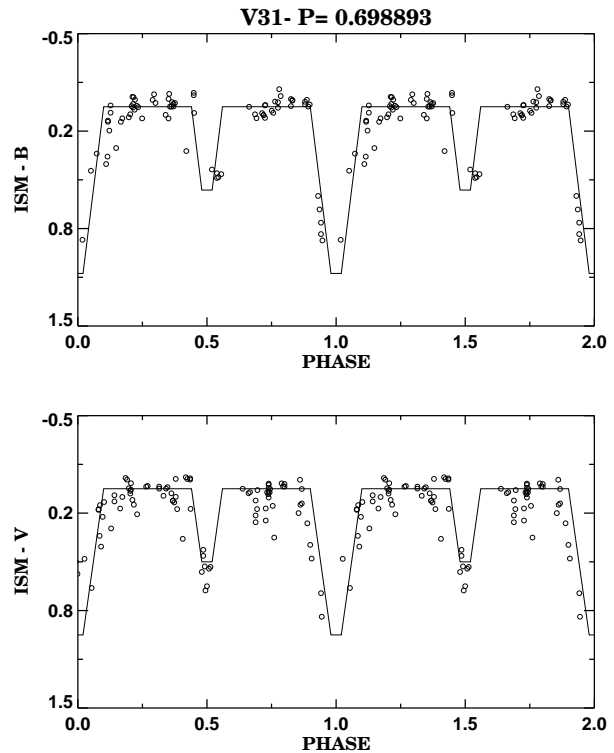, height=4.0in, width=8.5cm}}
      \caption{ISM light curves for V31, obtained using 
      {\sc isis}. Superimposed on the data is a light curve 
      `template' for Algol variables (from Layden \& Sarajedini 
      2000).}
      \label{Fig13}
\end{figure}

This having been stated, we emphasize that the light 
curve (Fig.~\ref{Fig13}) 
does {\em not} obviously resemble either those of BL~Her or of 
AHB1 stars (cf. Diethelm 1983). In fact, the light curve remarkably 
resembles that of an eclipsing binary.    
V31 might be a contact binary (W~UMa star). We have compared the 
properties of V31 
against those of W~UMa-type binaries in GCs from Rucinski (2000). 
Judging from Rucinski's eq.~(1) (a period-colour-luminosity relation), 
the period and colour of the star are inconsistent with its 
classification as a cluster W~UMa star: V31's implied absolute 
magnitude, $M_V \approx 1.54$~mag, corresponds to an apparent $V$ 
magnitude of $V \approx 19.0$~mag, adopting a distance modulus 
$(m-M)_V = 17.46$~mag (Harris 1996). This is $\sim 1.6$~mag fainter 
than V31, which has $\langle V\rangle \simeq 17.4$~mag (cf. Table~3). 
Alternatively, the star might be a field W~UMa variable. However,  
V31 is very close to the cluster centre, making it an unlikely (but 
not impossible) field variable. In addition, the colour and amplitude 
of the star, compared against the results compiled by Rucinski,
appear to militate against the very classificaion of V31 as a W~UMa
star. In particular, it is clear from Rucinski's Table~2 that the 
bluest W~UMa stars (with colours similar to V31's) tend to have much  
smaller amplitudes than is the case with V31. The star also lies 
outside the domain of W~UMa variables as shown in Rucinski's Fig.~6
(period-colour plane). Therefore, we suggest that this star is 
instead a short-period detached or semi-detached (most likely an 
Algol, but possibly a $\beta$~Lyrae) foreground eclipsing binary. 
This classification is certainly consistent with the star's light 
curve, as shown in Fig.~\ref{Fig13}, where the unequally deep 
eclipses stand out, and there is an indication of a period during 
which the light may be constant; accordingly, the Algol `template' 
from Layden \& Sarajedini (2000), kindly provided by A.~Layden, 
seems to provide a good fit to the star's light curve 
(Fig.~\ref{Fig13}).

However, as in the case of V27, we consider the periods and light 
curves of V31 preliminary. Additional data are needed to determine 
its variability status and ephemerides with greater confidence.

\section{Summary}
In the present paper, we have presented the first extensive CCD 
investigation of the variable star population in the outer-halo 
GC NGC~6229. With the exception of five variable stars from the 
candidate list presented by BCSS97, we were able to derive 
periods and $B,\,V$ light curves for 36 previously 
known/suspected variables in the field of the cluster, 
as well as for 12 newly discovered variables. Most of the new 
discoveries we owe to an application of the new ISM method 
developed by Alard (1999) and Alard \& Lupton (1998), which 
does appear to be a very powerful tool to search for new 
variables and determine their periods, especially in crowded 
regions such as the cores of GCs.  

The confirmed variable-star population in NGC~6229 is mostly 
comprised of RRab stars, several of which present the Blazhko 
effect. Less numerous, but also well represented, are the 
RRc stars. The mean periods of the fundamental RR Lyrae 
pulsators, as well as their relative incidence in comparison 
with first-overtone RR Lyrae variables, clearly classify 
NGC~6229 as an OoI cluster, as originally suggested by 
Mayer (1961). 

We have found that the period-amplitude distribution of 
NGC~6229 RRab variables with `regular' light curves may 
deviate from that of M3, a very well-known OoI 
globular. The period shift is in the sense that NGC~6229 
variables have shorter periods than their M3 counterparts at 
a given amplitude. Given that NGC~6229 appears to have 
the same metallicity as M5 (Borissova et al. 1999), which in 
turn is well known to be $\simeq 0.2-0.3$~dex more metal-rich 
than M3, this might indicate that the period-amplitude diagram 
is not solely a function of Oosterhoff type as recently suggested, 
but also of metallicity---a conclusion also reached by Kaluzny 
et al. (2000) in the case of M5. However, we caution that the size 
of the period shift was not consistent when $B$ or $V$ amplitudes
were employed, so that systematic effects cannot be 
ruled out at this point. Better data will be necessary before 
this problem can be conclusively solved. 

We have employed Fourier decompositions of the RRc and 
RRab light curves in an effort to determine their physical 
parameters (Simon \& Clement 1993; Kov\'acs \& Jurcsik
1996, 1997); the results were compared with those for other 
clusters (as published in the literature), and placed in the 
context of our period-shift discussion.  

Finally, we have found that NGC~6229 may contain up to two 
Pop.~II Cepheids of the W~Vir type, as well as an eclipsing 
binary (most likely a foreground Algol system) and an 
additional bright variable for which we were unable to 
reliably determine the variability type.

\section*{Acknowledgements}

The authors are grateful to C. Alard, B.W. Carney, C. Clement, 
J. Jurcsik, L.K. Fullton, A.C. Layden, A. Olech and R.T. Rood for 
very helpful discussions, information, and/or kindly supplied data. 
This research was supported in 
part by the Bulgarian National Science Foundation grant with the 
Bulgarian Ministry of Education and Sciences. Support for M.C. was 
provided by NASA through Hubble Fellowship grant HF--01105.01--98A 
awarded by the Space Telescope Science Institute, which is operated 
by the Association of Universities for Research in Astronomy, Inc., 
for NASA under contract NAS~5-26555.

\end{document}